\documentclass[prx,twocolumn,english,superscriptaddress,floatfix,longbibliography
]{revtex4-2}

\usepackage{graphicx}
\usepackage{dcolumn}
\usepackage{bm}
\usepackage{geometry}
\usepackage{braket}
\usepackage{relsize}
\usepackage{amsmath}
\usepackage[section]{placeins}
\usepackage{multirow}
\usepackage{tikz}
\newcommand{\etal}{\textit{et al.}}

\begin{document}

\preprint{AIP/123-QED}

\title{Energy Landscapes for the Quantum Approximate Optimisation Algorithm}

\author{Choy Boy}
\author{David J. Wales}
\affiliation{Yusuf Hamied Department of Chemistry, University of Cambridge, Lensfield Road, Cambridge, CB2 1EW, United Kingdom}

\begin{abstract}
Variational quantum algorithms (VQAs) have demonstrated considerable potential
in solving NP-hard combinatorial problems in the contemporary near
intermediate-scale quantum (NISQ) era. 
The quantum approximate optimisation algorithm (QAOA) is one such algorithm,
used in solving the maximum cut (Max-Cut) problem for a given graph by successive implementation of $L$ quantum circuit layers within a 
corresponding Trotterised ansatz.
The challenge of exploring the cost function of VQAs arising from an
exponential proliferation of local minima with increasing circuit depth has been well-documented.
However, fewer studies have investigated the impact of circuit depth on QAOA performance in finding the correct Max-Cut solution.
Here, we employ basin-hopping global optimisation methods to
navigate the energy landscapes for QAOA ans\"atze for various graphs, 
and analyse QAOA performance in finding the correct Max-Cut solution.
The structure of the solution space is also investigated using discrete path
sampling to build databases of local minima and the transition states that connect them, providing insightful visualisations using disconnectivity graphs.   
We find that the corresponding landscapes generally have a single funnel organisation,
which makes it relatively straightforward to locate low-lying minima with
good Max-Cut solution probabilities.
In some cases below the adiabatic limit the second lowest
local minimum may even yield a higher solution probability than the global
minimum. This important observation has motivated us to develop broader metrics
in evaluating QAOA performance, based on collections of minima obtained from
basin-hopping global optimisation.
Hence we establish expectation thresholds in elucidating useful solution probabilities from local minima,
an approach that may provide significant gains in elucidating reasonable solution probabilities from local minima.

\end{abstract}

\maketitle

\section{Introduction}
The initial setback of implementing practical quantum algorithms utilising the
quantum phase estimation (QPE) architecture onto current-day near
intermediate-scale quantum (NISQ) devices \cite{Malley2016, Leymann2020,
Skosana2021}, which typically possess short decoherence times
\cite{Burnett2019} and significant quantum noise \cite{Chen2023}, has prompted
the rapid development of variational quantum algorithms (VQAs) with
shorter quantum circuit depths
\cite{McClean2016,Peruzzo2014,Bharti2022}. VQAs typically operate within a
hybrid classical-quantum optimisation framework \cite{Yuan2019}, where an
initial quantum state is evolved by a parameterised circuit ansatz on a quantum
device. After the final evolved wavefunction is measured, a classical optimiser
evaluates the cost function from the measurement and subsequently suggests new
parameters that are fed back into the parameterised quantum circuit. This
interface between classical and quantum computers iterates until a
suitable convergence criterion is attained. VQAs have proved to be surprisingly
robust in tackling various sources of noise attributed to NISQ devices, such as
decoherence \cite{Ravi2022} and depolarisation \cite{Ito2023}. These properties are ascribed to
the innate variational nature acting as a parametric lever, which can be
flexibly adjusted even under noisy environments \cite{Fontana2021}. Coupled
with the recent advances in error-mitigation methods \cite{Halder2023,
Weidinger2023, Bravyi2022}, it is anticipated that VQAs will 
enable the realisation of practical quantum advantage before
the advent of fault-tolerant devices equipped with error-correction protocols \cite{Kim2023}.

The quantum approximate optimisation algorithm (QAOA) is a VQA that 
monotonically improves in performance as the number of quantum circuit
layers in the ansatz increases \cite{Farhi2014}. 
Variational quantum eigensolver (VQE) algorithms typically utilise the global minimisation of the cost landscape directly as the solution of interest
\cite{Zhang2022}. In contrast, QAOA uses the resulting final wavefunction to obtain
approximate solutions based on the states in the computational basis with the
highest frequency when measurement after circuit evolution is carried out. Hence
QAOA is a particularly attractive algorithm for solving combinatorial optimisation
problems, such as Max-Cut \cite{Wurtz2022}, with 
promising applications in portfolio optimisation \cite{Buonaiuto2023} and
chemistry \cite{Kremenetski2021,Hadfield2022}.

The expansion of VQAs has also motivated the study of their potential
shortcomings in solving practical large-scale problems 
from a software perspective. In particular, for various VQA ans\"atze there is an exponential growth in the
barren plateau problem as the number of qubits and circuit layers required to
encode a given problem increases 
\cite{Wang2021,McClean2018,Cerezo2021}. Recently, more insight has also been
gained into the challenges of exploring the cost landscapes of VQAs that arise
from a proliferation in the number of local minima and other stationary points
as the complexity of the problem increases \cite{Wierichs2020,Anschuetz2022}.
However, further analysis of the organisation of the global cost landscapes of
VQAs, and how this structure impacts the quality of the solutions obtained, is
needed, especially for QAOA \cite{Sack2023, Crooks2018}.

Here, we seek to address these gaps in understanding using the well-established theory and associated
computational methodology of molecular energy landscapes \cite{Wales03,Wales18}.
We characterise the cost optimisation landscapes of QAOA for
various weighted and unweighted graphs in solving the Max-Cut problem 
using basin-hopping global optimisation methods \cite{lis87,walesd97a,waless99} to locate global minima, and discrete path sampling \cite{Wales02,Wales04} to
create connected databases of minima and the transition states that connect them.
Recently, energy landscape techniques have demonstrated considerable utility for
quantum computing in the analysis of hardware-efficient ans\"atze for the VQE algorithm \cite{Choy2023}, and optimisation of
electronic wavefunctions in a combined discrete space of operators and continuous
parameter amplitudes \cite{Burton2023}. Our new results for QAOA show that the
solution landscapes below the adiabatic limit generally possess single-funnelled
structures associated with self-organising systems where locating the global minimum is relatively straightforward \cite{walesmw98,Wales03,BogdanW04}.
Furthermore, we find that local minima sufficiently close in
energy to the global minimum may also exhibit good solution probabilities
for the Max-Cut problem. In some instances, the second lowest minimum has
a higher solution probability than the global minimum, highlighting the
importance of studying the VQA solution landscape globally. This observation 
leads us to introduce metrics that take into account the distribution of
minima in evaluating the performance and robustness of QAOA. We also utilise
the convex hull of the solution space in estimating expectation cutoffs for the
location of local minima with reasonable solution probabilities. We hope
that these techniques can advance the feasibility of implementing QAOA for
problems with numerous local minima in noisy environments.

\section{Methodology}
Given an undirected graph $G=(V,E)$, with weights $w_{ij}$ assigned to edges
$e_{ij}\in{E}$ for connected vertices $i,j\in{V}$, the Max-Cut problem seeks to
partition $V$ into two distinct sets such that the sum of weights between the
two sets is maximised. If $w_{ij} = 1$ for all $e_{ij}$, then $G$ is said to be
an unweighted graph; otherwise $G$ is a weighted graph. It follows that the
Max-Cut problem can be mapped to a two-spin Ising-type cost Hamiltonian
$\hat{H}_C$ corresponding to $N$ implementable qubits:
\begin{equation}
\hat{H}_C = \frac{1}{2}\mathlarger{\sum}_{e_{ij}\in{E}}{w_{ij}}(Z_{i} \otimes Z_{j}),
\end{equation}
where the states $\ket{s} = \{\ket{\alpha}, \ket{\beta}\}^{\otimes{N}}$ encode
the desired solution strings to the Max-Cut problem, with
$\ket{\alpha}=\ket{0}$ if and only if $\ket{\beta}=\ket{1}$, and vice versa.
Thus, the aim of QAOA is to approximate the ground-state energy or the lowest
eigenvalue of $\hat{H}_C$ via a suitable ansatz with unitary operator
$\hat{U}(\bm{\theta})$ to evolve an initial state $\ket{\psi_{0}}$, and
subsequently use the final evolved state
$\ket{\Psi(\bm{\theta})}=\hat{U}(\bm{\theta})\ket{\psi_{0}}$ to approximate
$\ket{s}$. This objective can be achieved on a quantum device by performing a
certain number of shots per experiment and measuring all qubits in the
computational basis, taking the state possessing the greatest number of shots
to best approximate $\ket{s}$ for that experiment. We seek to simulate this
procedure classically by considering the probability of measuring the state
$\ket{s}$ in the computational basis, $p(\ket{s})$:
\begin{equation}
p(\ket{s}) = |\braket{s|\Psi(\bm{\theta})}|^{2}.
\end{equation}
The objective function to be minimised by the classical computer is the expectation of $\hat{H}_C$, $\langle \hat{H}_C \rangle$:
\begin{equation}
\langle \hat{H}_C \rangle = {E({\boldsymbol{\theta}})} = {\bra{\psi_0}\hat{U}^{\dag}(\bm{\theta})\hat{H}_{C}\hat{U}(\bm{\theta})\ket{\psi_0}}.
\end{equation}
The QAOA ansatz with parameters $\bm{\theta}=\{\bm{\gamma},\bm{\delta}\}$ can be assembled as a Trotterised variational schedule, comprising 
a cost circuit layer with unitary operator $\hat{U}_C(\gamma)$, followed by a mixer circuit layer with unitary operator $\hat{U}_M(\delta)$ up to a circuit depth $L$:
\begin{equation}
\ket{\Psi(\bm{\gamma},\bm{\delta})}=\mathlarger{\prod_{l=1}^{L}}\hat{U}_M(\delta_l)\hat{U}_C(\gamma_l)\ket{\psi_0},
\end{equation}
where $\ket{\psi_0}=\ket{+}^{\otimes{N}}$ is the state encoding for all possible partitions of $V$ with equal probability. The cost layer encapsulating $\hat{H}_C$ can be compiled as a sequence of two-qubit parameterised $R_{zz}$ quantum gates for qubits $q_i$ and $q_j$, with $\gamma$ scaled based on the weights of $e_{ij}$:
\begin{equation}
\begin{split}
\hat{U}_C(\gamma) &=e^{-i\gamma\hat{H}_{C}} \\
&=\mathlarger{\prod}_{e_{ij}\in{E}}R_{zz}(-w_{ij}\gamma).
\end{split}
\end{equation}
The mixer layer performs a time-evolution of the mixer Hamiltonian $\hat{H}_{M}=-\sum_{i=1}^{N}{X_i}$, which 
anticommutes with $\hat{H}_{C}$ and has $\ket{\psi_0}$ as an eigenvector. The mixer layer can be realised as a parallelisation of single-qubit parameterised $R_x$ quantum gates:
\begin{equation}
\begin{split}
\hat{U}_M(\delta) &= e^{-i\delta\hat{H}_{M}} \\
&=\mathlarger{\bigotimes}_{i=1}^{N}{R_x}(2\delta).
\end{split}
\end{equation}
It has been shown that QAOA conforms to the adiabatic theorem, i.e.~for
$L\rightarrow\infty$ the final evolved state $\ket{\Psi(\bm{\theta})}$
converges exactly to the ground state of $\hat{H}_{C}$, and thus gives the
optimal $p(\ket{s})$ \cite{Farhi2014}. In practice, such an implementation is
unfeasible in the NISQ regime, hence we are interested in considering $L_{ad}$
for a given system, defined as the minimum number of circuit layers required to reach
the adiabatic limit, assuming that $L_{ad}$ can be attained. As we will
demonstrate in our analysis of the energy landscapes of QAOA, it is also
important to distinguish $L_{ad}$ from $L_{min}$, where $L_{ad} \geq L_{min}$.
Here, $L_{min}$ is the minimum number of layers needed to achieve the maximum
$p(\ket{s})$ in the corresponding global minimum. Local
minima with lower $p(\ket{s})$ may be present due to underparametrisation
of the circuit ansatz, and hence a less thorough exploration of states in the
Hilbert space may be sufficient to obtain a useful solution. We hypothesise that the exponential
increase in the number of local minima is attributable to circuit ans\"atze
with $1\leq{L}\leq{L_{min}}$ layers. The behaviour of local minima may vary for
$L_{min}<{L}\leq{L_{ad}}$ layers if $L_{min}<{L_{ad}}$, and we observe for
various graphs that the number of local minima may increase first before
decreasing to the adiabatic limit, or instead decrease monotonically.

For each graph considered and $L$, we generate an initial
set of minima via basin-hopping global optimisation \cite{lis87,walesd97a,waless99} using the GMIN
program \cite{GMIN}. The analytic gradients of the parameterised rotation
gates were calculated via the parameter-shift rule \cite{Mari2021}
(see \textbf{Appendix A} for more details):
\begin{equation}
\frac{\partial {E({\boldsymbol{\theta}})}}{\partial \theta_i} = \frac{1}{2}\left[ {E\left({\boldsymbol{\theta}}+\frac{\pi}{2}\bm{e}_i\right)-E\left({\boldsymbol{\theta}}-\frac{\pi}{2}\bm{e}_i\right)} \right].
\end{equation} 
Local minimisation for the basin-hopping steps employed a limited-memory Broyden \cite{Broyden70}, Fletcher 
\cite{Fletcher70}, Goldfarb \cite{Goldfarb70}, Shanno \cite{Shanno70}
(L-BFGS) procedure \cite{lbfgs,Nocedal80} equipped with the Metropolis criterion for
accepting/rejecting steps \cite{Paleico2020}.
The resulting minima were then employed as the starting points for construction of a kinetic transition
network \cite{NoeF08,pradag09,Wales10a}.
Discrete path sampling  \cite{Wales02,Wales04} (DPS) was used via connection attempts for selected
minima pairs with final
states $\ket{\Psi(\bm{\theta}_{\mu})}$ and $\ket{\Psi(\bm{\theta}_{\nu})}$.
The doubly-nudged \cite{TrygubenkoW04,SheppardTH08} elastic band \cite{Millsjs95,JonssonMJ98,HenkelmanUJ00,HenkelmanJ00}
approach was used to locate candidates for accurate transition state refinement by
hybrid eigenvector-following \cite{munrow99,HenkelmanJ99,KumedaMW01}.
The missing connection algorithm \cite{CarrTW05} was used to select pairs of minima to fill in the
gaps in incomplete pathways via Dijkstra's shortest path algorithm \cite{Dijkstra59} combined with
a distance metric based on the state overlap between local minima, $S_{\mu\nu}$:
\begin{equation}
S_{\mu\nu} = 1 - |\braket{{\Psi(\bm{\theta}_{\mu})}|{\Psi(\bm{\theta}_{\nu})}}|.
\end{equation} 
Any new minima are added to the database along with the transition states and connection information.
The resulting cost landscapes can be visualised using disconnectivity graphs,
where the local minima are segregated into disjoint sets for regular thresholds in the energy \cite{BeckerK97,walesmw98}.
In these graphs, the bottom of a branch corresponds to the energy of a local minimum, and the branches are
joined when the corresponding minima can interconvert via a pathway below the given threshold.
Visualisation of the energy landscape can be further enhanced by colouring each minimum with the
corresponding probability of finding the correct Max-Cut solution; we find this construction especially useful in 
comparing the solution landscapes as $L$ varies.

\begin{figure}[ht]
\includegraphics[width=8.2cm]{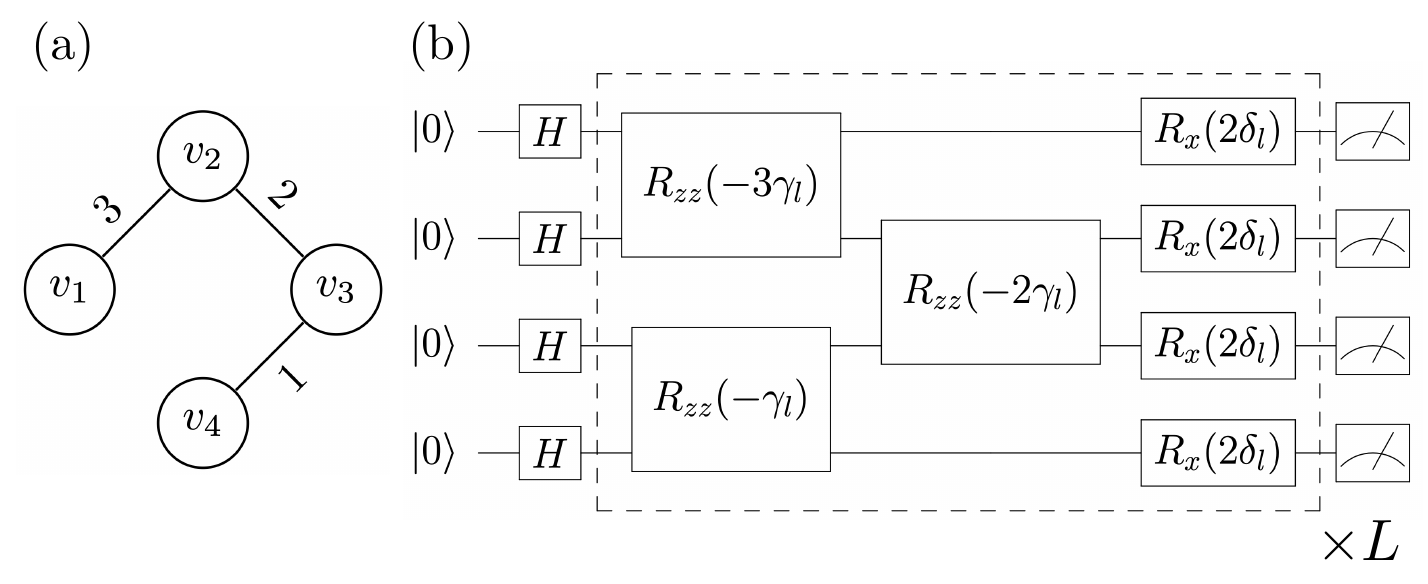}
\caption{\label{fig:1} \textbf{(a) (left)} Weighted graph $G_1$ encoded with four qubits. \textbf{(b) (right)} Corresponding QAOA ansatz for $G_1$.}
\end{figure}
\begin{figure}[ht]
\includegraphics[width=8.2cm]{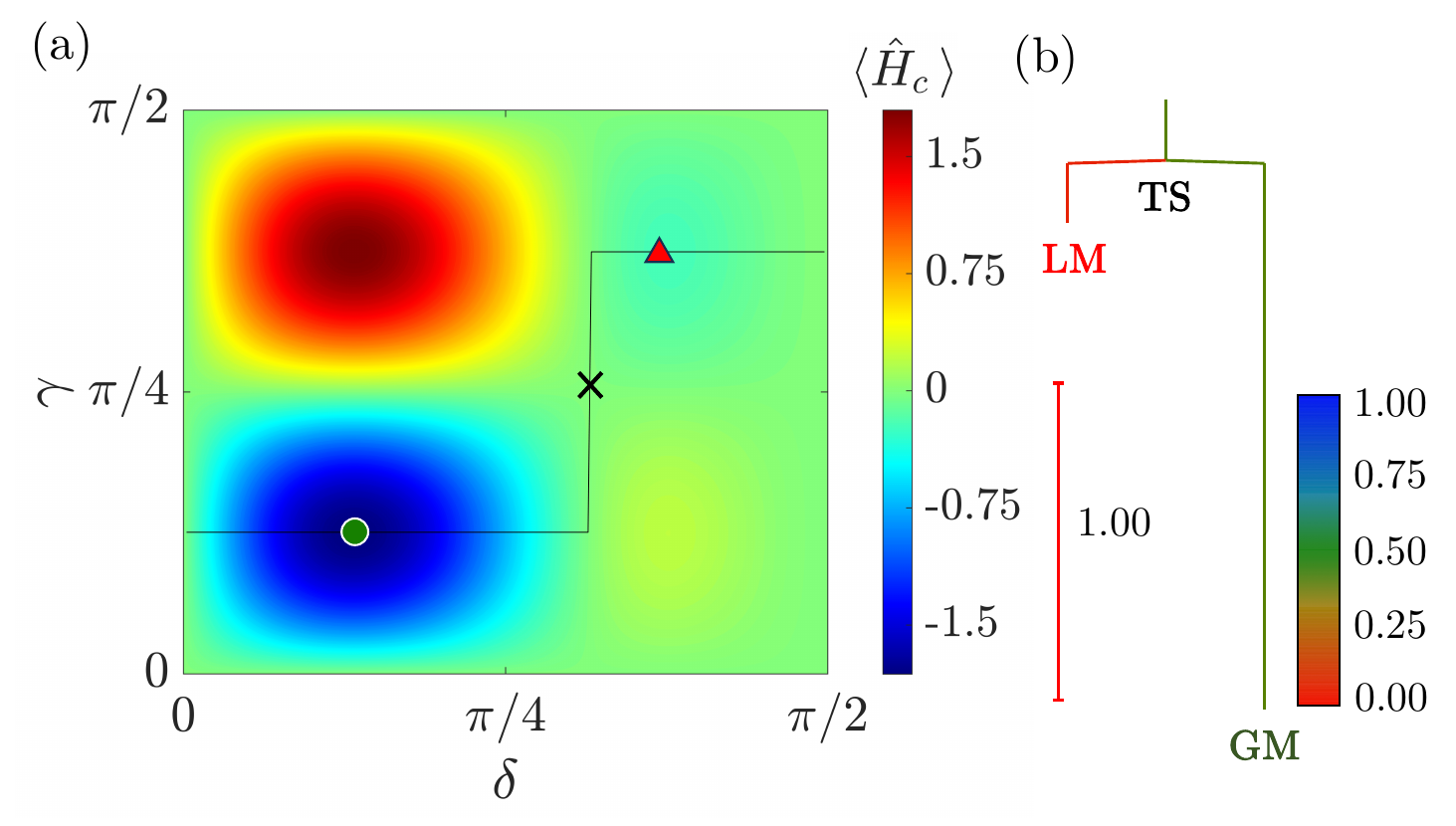}
\caption{\label{fig:2} \textbf{(a) (left)} Contour plot of $\langle \hat{H}_C
\rangle$ for graph $G_1$ and its QAOA ansatz with $L=1$ against the parameters
$\gamma$ and $\delta$. The solid line depicts the minimum of each
vertical slice of the contour plot, with the global minimum (\textbf{GM}, green circle), local minimum (\textbf{LM}, red triangle) and the transition state connecting
them (\textbf{TS}, black cross) situated on the pathway that is also plotted. \textbf{(b)
(right)} Corresponding disconnectivity graph of the contour plot in
\textbf{Fig. 2a}, with both the \textbf{GM} and \textbf{LM} coloured (grayscaled) based on
their respective probabilities of obtaining the correct Max-Cut solution of
$\ket{\alpha\beta\alpha\beta}$.}
\end{figure}

As an example, consider the weighted graph $G_1$ with four vertices
(\textbf{Fig.~1a}), where the Max-Cut problem can be encoded as a four-qubit
QAOA ansatz with varying $L$ (\textbf{Fig.~1b}). We note that
although the ansatz cost circuit layer can be compiled in numerous ways, the
arrangement in \textbf{Fig.~1b} is ideal in reducing the overall circuit depth.
This ordering does not require swap gates to permute
non-neighbouring qubits corresponding to their respective edges, which is an important
consideration when transpiling QAOA onto planar superconducting processors
\cite{Harrigan2021}. The cost landscape of the $L=1$ circuit ansatz features a
global and a local minimum connected by a transition state (\textbf{Fig.~2a}), and the corresponding disconnectivity
graph is shown in \textbf{Fig.~2b}, where the branches are coloured with the probabilities of finding the state
$\ket{\alpha\beta\alpha\beta}$, which corresponds to the Max-Cut solution of $G_1$.

\section{Results}
\subsection{Complete unweighted graphs}
We begin by examining the QAOA circuit ans\"atze for the complete unweighted
graph series $K_N$ from $N=3$ to $N=8$ and from $L=1$ to $L=3$, where each
vertex is connected to every other vertex with unit weight. It follows that
the Max-Cut solution of $K_N$ is the set of all possible tensor product
permutations of $\lfloor N/2 \rfloor$ number of $\ket{\alpha}$ states and
$\lceil N/2 \rceil$ number of $\ket{\beta}$ states: thus the total number of
Max-Cut solutions for odd $N$ is $2N!/\{\lfloor N/2 \rfloor! \lceil N/2 \rceil!
\}$, and for even $N$ $N!/\{2(N/2)!\}$ solutions.
\begin{table}
\caption{\label{tab:1} Number of minima $M$ (top value) and the highest correct
Max-Cut probability (HCMP) (bottom value) for graphs $K_3$ to $K_8$ with varying
$L$ obtained from basin-hopping global optimisation. For the case of $K_8$ and $L=3$, the
HCMP is not equal to the maximum value of 1, and the number of decimal
places used for all HCMPs is chosen to be the same as this case for ease of comparison.}
\begin{ruledtabular}
\begin{tabular}{cccc}
\textbf{Graph} & $L=1$ & $L=2$ & $L=3$ \\
\hline
\multirow{2}{*}{\centering $K_3$}
      & 1    & 1    & 1    \\
      & 1.000000    & 1.000000    & 1.000000    \\
\hline
\multirow{2}{*}{\centering $K_4$}
      & 1    & 1    & 1    \\
      & 0.739106    & 1.000000    & 1.000000    \\
\hline
\multirow{2}{*}{\centering $K_5$}
      & 1    & 4    & 1    \\
      & 0.975990    & 1.000000    & 1.000000    \\
\hline
\multirow{2}{*}{\centering $K_6$}
      & 1    & 23    & 324    \\
      & 0.671340    & 0.994239    & 1.000000    \\
\hline
\multirow{2}{*}{\centering $K_7$}
      & 1    & 37    & 598    \\
      & 0.951350    & 0.999619    & 1.000000    \\
\hline
\multirow{2}{*}{\centering $K_8$}
      & 1    & 46    & 3418    \\
      & 0.629727    & 0.991483    & 0.999997    \\
\end{tabular}
\end{ruledtabular}
\end{table}
\begin{figure*}[t]
\centering
\includegraphics[width=17cm]{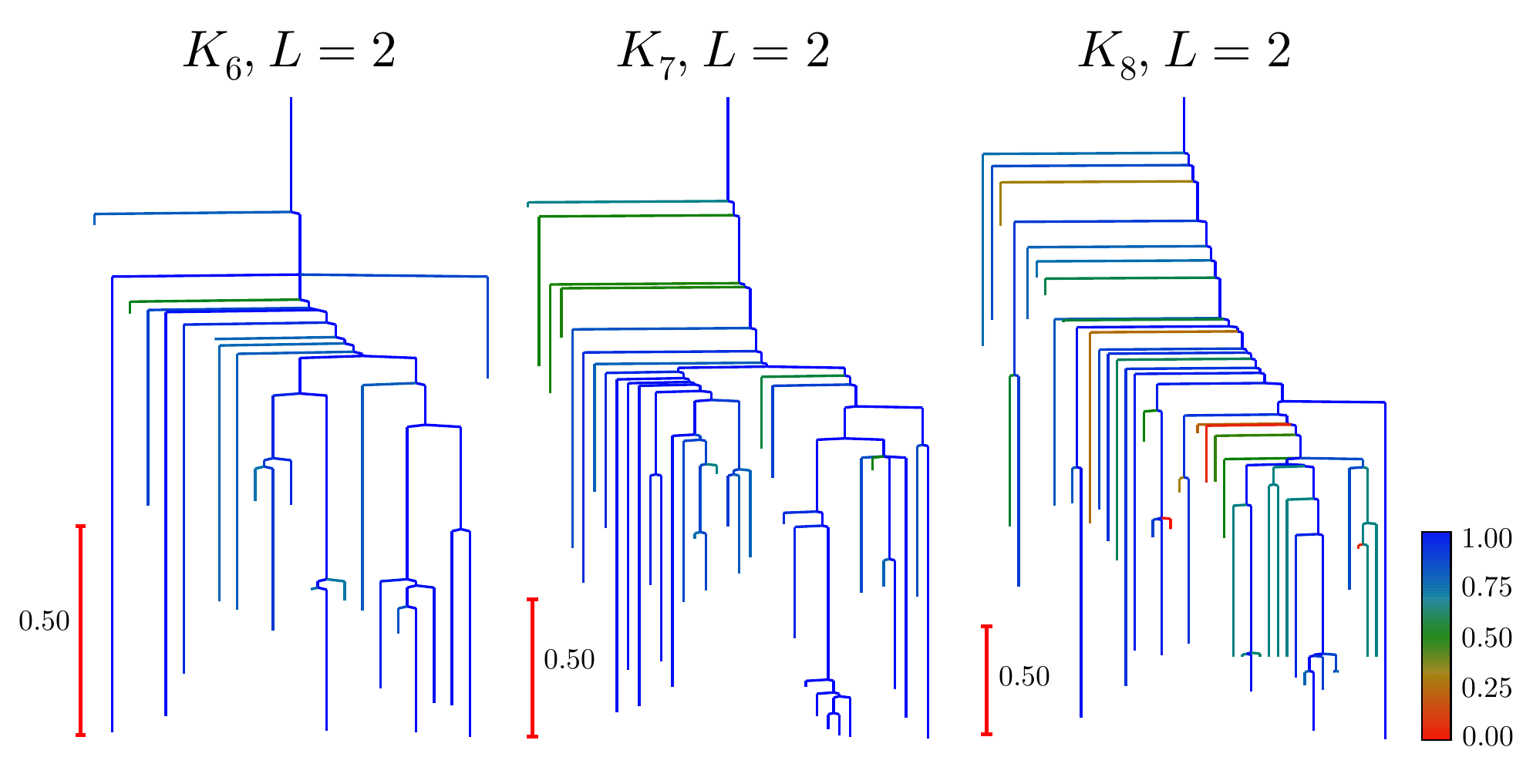}
\caption{\label{fig:3}Disconnectivity graphs of $K_6$, $K_7$ and $K_8$ for $L=2$.}
\vspace{0.4cm}
\includegraphics[width=17cm]{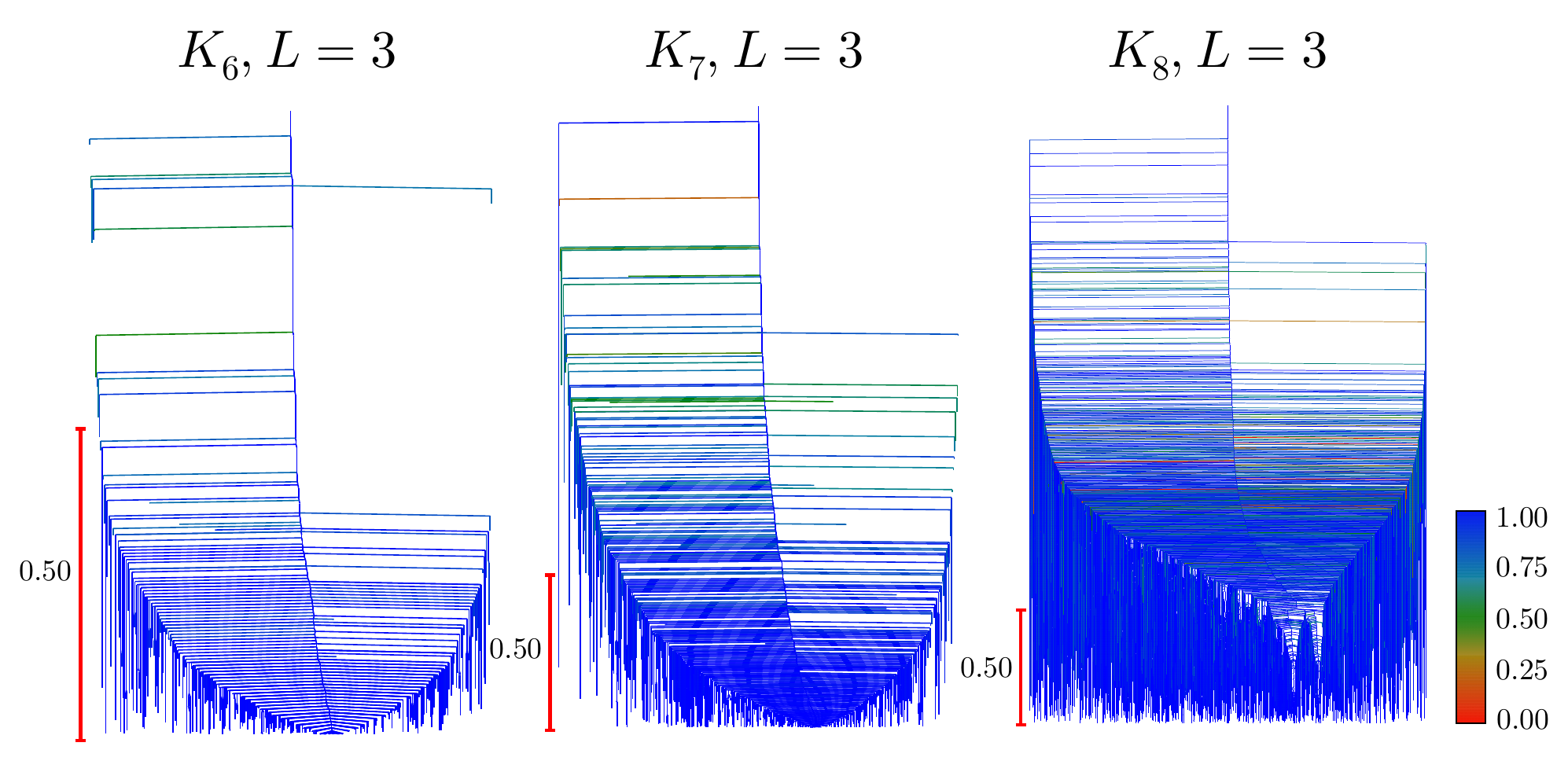}
\caption{\label{fig:4}Disconnectivity graphs of $K_6$, $K_7$ and $K_8$ for $L=3$.}
\end{figure*}

\textbf{Table I} summarises the number of minima $M$ and the highest correct
Max-Cut probability (HCMP) from the collections of minima obtained via
basin-hopping global optimisation for $L=1$ to $L=3$. The complete graphs with
odd $N$ generally possess higher HCMPs than their even counterparts,
mainly due to their greater number of accepted Max-Cut solutions that
contribute to their corresponding probabilities. We also find that although for
$K_6$ to $K_8$ the expected exponential increase in $M$ is observed as $L$
increases, for $K_5$ there was a decrease in the number of minima from $L=2$ to
$L=3$, leading to a simplification in the energy landscape from $L_{min}=2$ to
$L_{ad}=3$. Looking at the disconnectivity graphs of $K_6$ to $K_8$ for $L=2$
(\textbf{Fig. 3}) and $L=3$ (\textbf{Fig. 4}), we observe that the majority of
the local minima generally possess very high correct Max-Cut probabilities,
particularly those closer to the global minimum, especially as $L$ and $M$
increase. The well-funnelled organisation of the landscape also becomes more
apparent as $L$ increases, and this structure is expected to outweigh the
challenges associated with solving the Max-Cut problem for higher $N$,
particularly for $K_8$, where local minima with low probabilities are
increasingly interspersed with other local minima corresponding to higher
probabilities.
\begin{figure*}[t]
\centering
\includegraphics[width=17cm]{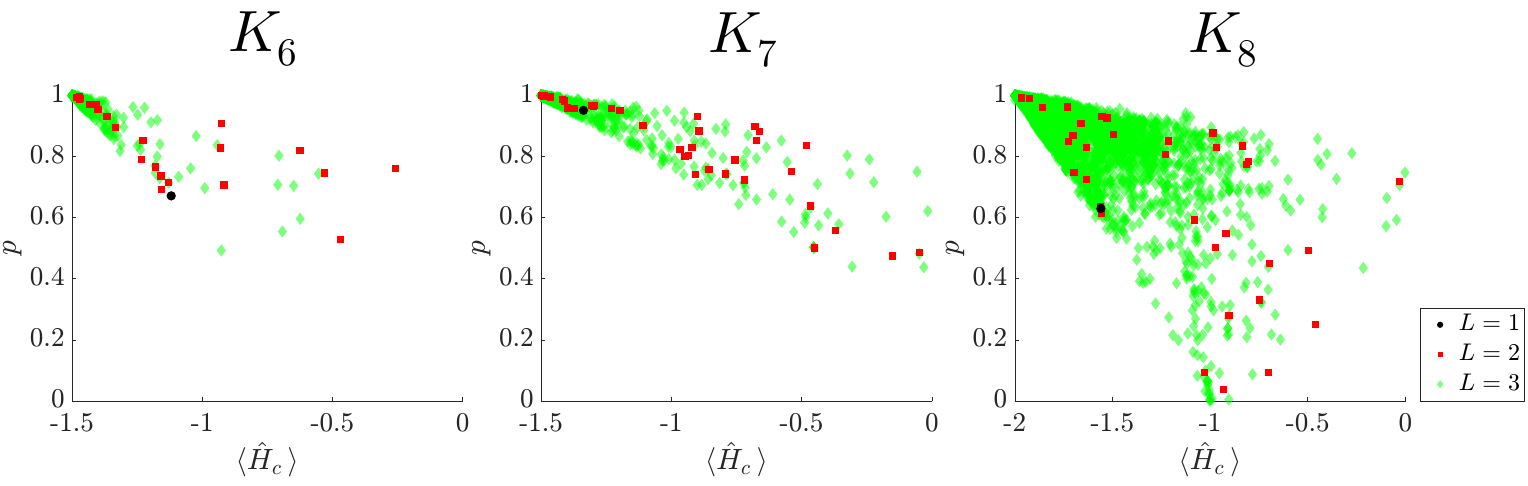}
\caption{\label{fig:5}Scatter plots of probabilities of the correct Max-Cut
solutions against $\langle \hat{H}_C \rangle$ for individual minima of graphs
$K_6$, $K_7$ and $K_8$ for $L=1$ (black circle), $L=2$ (red square) and $L=3$ (green diamond).}
\end{figure*}

To further evaluate the performance of QAOA ans\"atze for various $L$ based on
the databases of minima and their respective Max-Cut probabilities and
$\langle \hat{H}_C \rangle$ values, we introduce the weighted average metric $F$:
\begin{equation}
F = \frac{1}{M|{\langle \hat{H}_C \rangle}_{min}|}\mathlarger{\sum_{m=1}^{M}}|{\langle \hat{H}_C \rangle}_{min} - {\langle \hat{H}_C \rangle}_{m}|[1-p_m(\ket{s})],
\end{equation}
where ${\langle \hat{H}_C \rangle}_{min}$ is the ${\langle \hat{H}_C \rangle}$
value of the global minimum. This formulation of $F$ is advantageous for two
reasons. First, it distinguishes circuit ans\"atze with $L_{ad}$ and
$L_{min}$ layers, since for $L_{ad}$, $F=0$ because only the global minimum is
present, compared to $L_{min}$ where other local minima with lower Max-Cut
probabilities are present. Second, $F$ reflects the contribution of minima for
circuit ans\"atze with $L<L_{min}$ more accurately, since it is possible for
minima with lower ${\langle \hat{H}_C \rangle}$ values, including the global
minimum, to possess lower Max-Cut probabilities than their counterparts with
higher ${\langle \hat{H}_C \rangle}$ values. As $L$ increases, a decrease in
the value of $F$ can generally be interpreted as an improvement in QAOA
performance, since it corresponds to an increase in the proportion of local
minima with better probabilities. However, the converse situation, where $F$
increases as $L$ increases, may not necessarily signify a drop in QAOA
performance, as the well-funnelled organisation of the cost landscape, as well
as the guarantee of obtaining a better Max-Cut probability, may outweigh the
trade-off in obtaining a lower proportion of local minima with relatively good
probabilities. Nevertheless, we propose choosing circuit ans\"atze with $L$
layers that feature sufficiently low values of $F$ when simulating QAOA on
noisy quantum devices, as choosing circuits with higher $L$ may also increase
the impact of quantum gate and qubit decoherence noise. 
\begin{table}[ht]
\caption{\label{tab:2} The weighted average metric $F$ for graphs $K_5$ to $K_8$ of varying $L$.}
\begin{ruledtabular}
\begin{tabular}{cccc}
\textbf{Graph} & $L=1$ & $L=2$ & $L=3$ \\
\hline
$K_5$ & 0.002276 & 0.067219 & 0.000000 \\
\hline
$K_6$ & 0.083438 & 0.060651 & 0.007907 \\
\hline
$K_7$ & 0.005269 & 0.119087 & 0.026351 \\
\hline
$K_8$ & 0.081604 & 0.190701 & 0.038062 \\
\end{tabular}
\end{ruledtabular}
\end{table}

Analysing the $F$ values for $K_5$ to $K_8$ (\textbf{Table II}), we observe
that for $K_5$, $F=0$ for $L_{ad}=3$, differentiating it from $L_{min}=2$, as
expected. Interestingly, with the exception of $K_6$, $F$ appears to increase
for $L=2$ before decreasing for $L=3$. The increase in $F$ can mainly be
attributed to the general increase in the number of local minima for $L=2$ with
comparably lower probabilities than that of the single global minimum for
$L=1$. This trend is evident in the scatter plots of the probabilities of the
correct Max-Cut solution against $\langle \hat{H}_C \rangle$ for the databases
of minima for varying $L$ (\textbf{Fig. 5}). The somewhat triangular convex
hull of the solution space, $C_{s}$, seems to become better defined with the
transition from $L=2$ to $L=3$. We also see a proliferation of local
minima towards the apex of the global minimum, which would explain the observed
subsequent decrease in $F$. Thus, a choice of $L=3$ would be adequate for
solving the Max-Cut problem for graphs $K_5$ to $K_8$ based on their $F$
values.

\begin{figure}[ht]
\includegraphics[width=8.2cm]{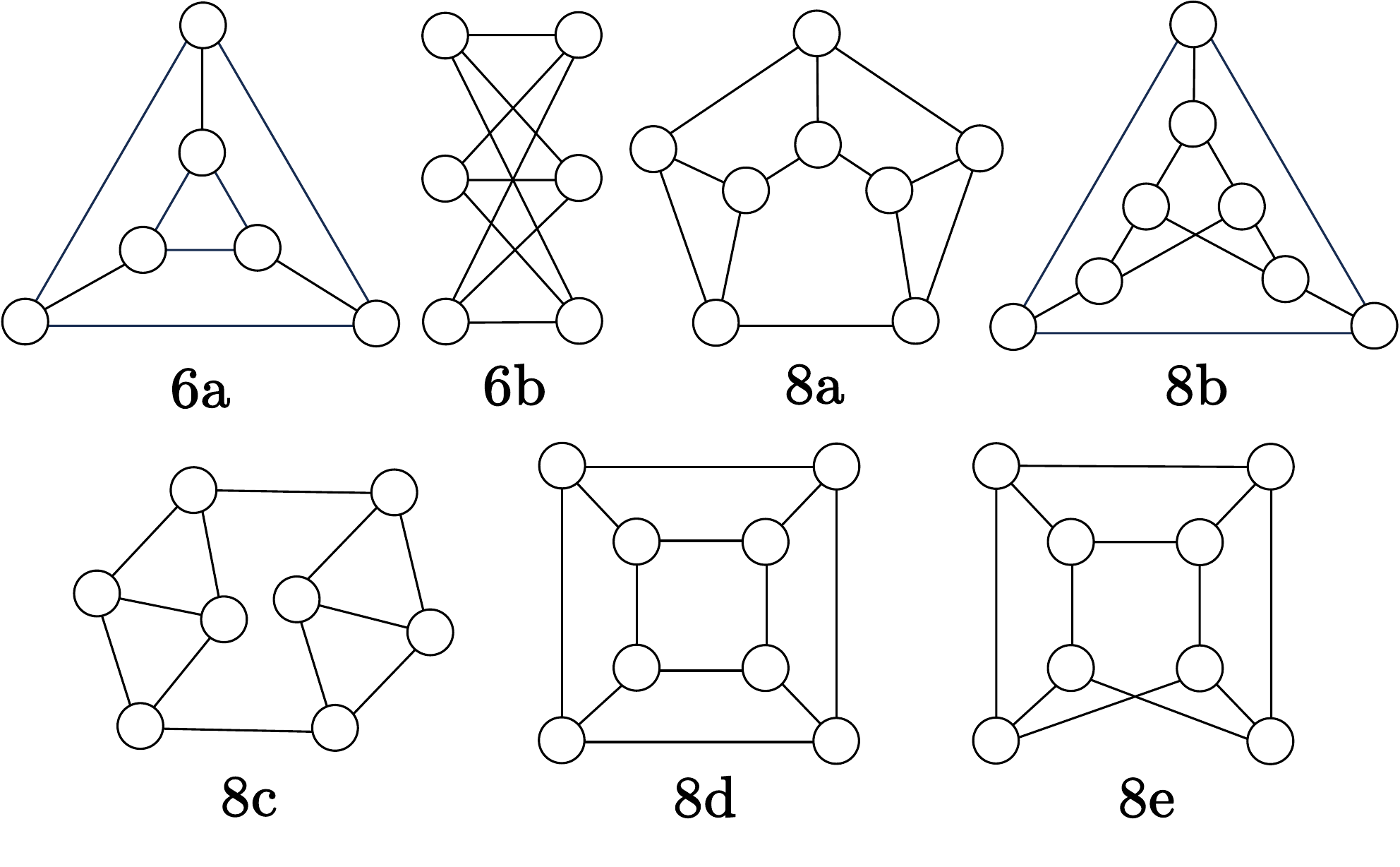}
\caption{\label{fig:6} 3-regular unweighted graphs \textbf{6a}–\textbf{8e} investigated in this study.}
\end{figure}
\begin{table}
\caption{\label{tab:3} Number of minima $M$ (top value), HCMPs (middle value)
and $F$ (bottom value) for graphs \textbf{6a}–\textbf{8e} of varying $L$
obtained from basin-hopping global optimisation. HCMPs with asterisks
indicate that they correspond to the next highest local minimum rather than
the global minimum.}
\begin{ruledtabular}
\begin{tabular}{ccccc}
\textbf{Graph} & $L=1$ & $L=2$ & $L=3$ & $L=4$ \\
\hline
\multirow{3}{*}{\centering \textbf{6a}}
      & 1    & 5    & 23 & 145   \\
      & 0.400816 & 0.720917 & 0.933445* & 0.996304   \\
      & 0.254240 & 0.224891 & 0.167129 & 0.093993   \\
\hline
\multirow{3}{*}{\centering \textbf{6b}}
      & 3    & 20    & 191 & 1451   \\
      & 0.274835 & 0.659823 & 0.915348 & 0.978625   \\
      & 0.772861 & 0.615867 & 0.510817 & 0.340108   \\
\hline
\multirow{3}{*}{\centering \textbf{8a}}
      & 1    & 4    & 16 & 83   \\
      & 0.142701 & 0.349371 & 0.616672 & 0.748746  \\
      & 0.427166 & 0.363476 & 0.497637 & 0.428079  \\
\hline
\multirow{3}{*}{\centering \textbf{8b}}
      & 1    & 4    & 15 & 97   \\
      & 0.232056 & 0.420045* & 0.638057 & 0.767138  \\
      & 0.355057 & 0.325065  & 0.426828 & 0.380643  \\
\hline
\multirow{3}{*}{\centering \textbf{8c}}
      & 1    & 4    & 18 & 82   \\
      & 0.077352 & 0.182753 & 0.536762 & 0.701246  \\
      & 0.516636 & 0.560946 & 0.519856 & 0.429701  \\
\hline
\multirow{3}{*}{\centering \textbf{8d}}
      & 1    & 6    & 31 & 151   \\
      & 0.186302 & 0.520680 & 0.871573 & 0.972013  \\
      & 0.500506 & 0.360078 & 0.258910 & 0.141219   \\
\hline
\multirow{3}{*}{\centering \textbf{8e}}
      & 1    & 4    & 15 & 110   \\
      & 0.321737 & 0.574918* & 0.769975 & 0.918878*  \\
      & 0.286668 & 0.241691  & 0.302882 & 0.300042   \\
\end{tabular}
\end{ruledtabular}
\end{table}
\begin{figure*}[t]
\centering
\includegraphics[width=17cm]{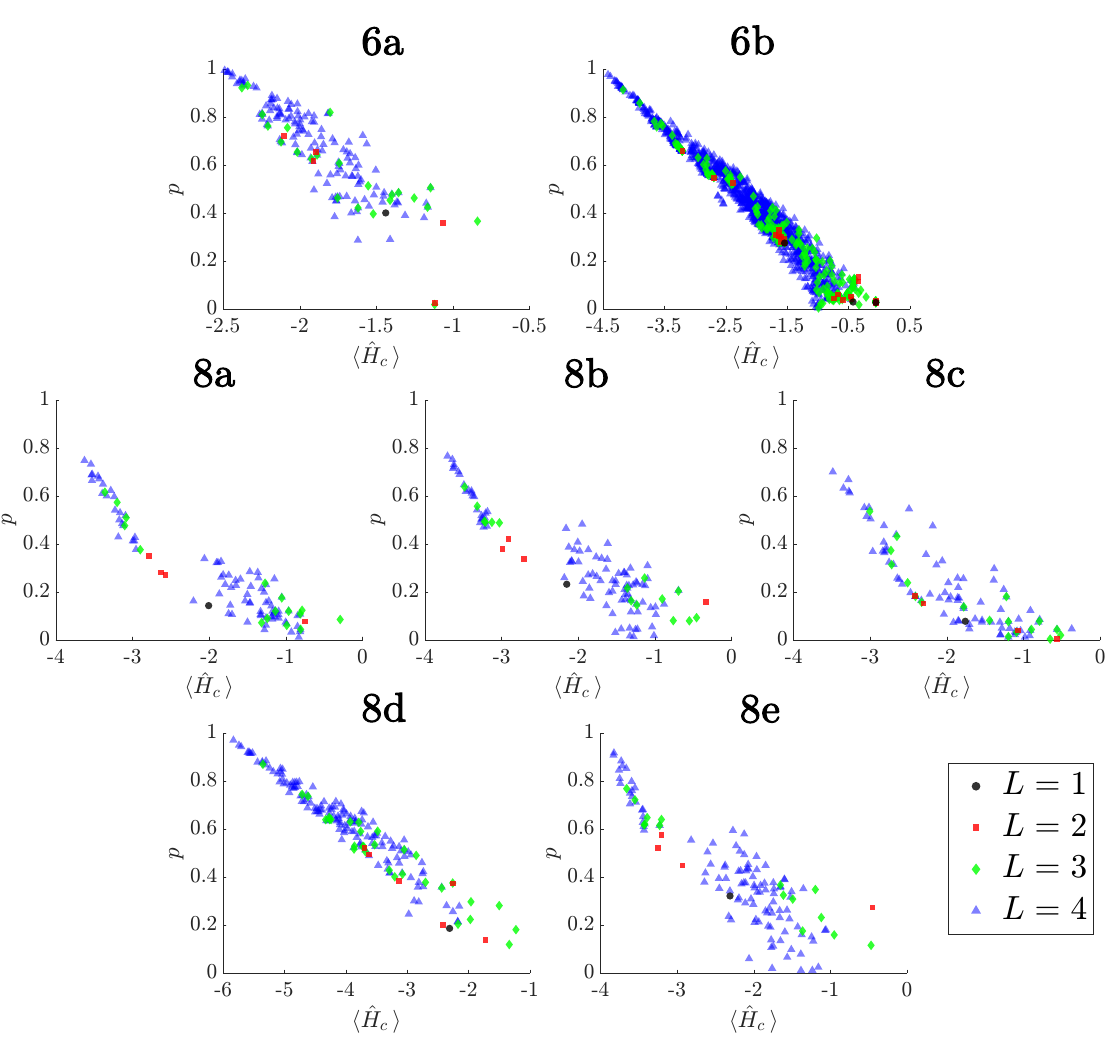}
\caption{\label{fig:7}Scatter plots of probabilities of the correct Max-Cut solutions against $\langle \hat{H}_C \rangle$ for connected minima of graphs \textbf{6a}–\textbf{8e} for $L=1$ (black circle), $L=2$ (red square), $L=3$ (green diamond) and $L=4$ (blue triangle).}
\end{figure*}
\subsection{3-regular unweighted graphs}
Next, we analysed all connected 3-regular unweighted graphs with six and eight
vertices \cite{Guerreschi2019} from $L=1$ to $L=4$, labelled
\textbf{6a}–\textbf{8e}, respectively (\textbf{Fig. 6}). In terms of the number of
minima obtained from basin-hopping runs, the 3-regular graphs
generally possess lower $M$ values than the complete graphs of $K_6$
and $K_8$, particularly for graphs with eight vertices (\textbf{Table III}).
For graph \textbf{6b}, starting from $L=1$ its QAOA ansatz gives rise to more
than one minimum, which subsequently produces a more rapid increase in $M$ 
compared to its counterpart \textbf{6a}. This phenomenon can largely be
attributed to the relatively more complex analytic expression of
$\langle \hat{H}_C \rangle$ with the $L=1$ circuit ansatz \cite{Pendas2022} for \textbf{6b}.
Another interesting pair of graphs is \textbf{8d} and \textbf{8e}, where,
although their $L=1$ analytic expressions of $\langle \hat{H}_C \rangle$ are
identical and thus give rise to the same $\langle \hat{H}_C \rangle_{min}$
value \cite{Pendas2022}, we find that the trends of $M$ and HCMPs for higher
$L$ are markedly different. Although the 3-regular series of
graphs give rise to lower $M$ for similar $L$, and hence somewhat simpler
energy landscapes compared to their complete graph counterparts, their HCMPs
are also comparatively lower, hence requiring a greater $L$ to achieve a
sufficiently high HCMP. In some cases, the HCMPs were derived not from the
global minimum, but from the next highest local minimum. (See \textbf{Appendix
B} detailing the differences in Max-Cut probabilities and $\langle \hat{H}_C
\rangle$ values between the two minima for these cases.) This phenomenon
appears sporadically without much predictability, most notably for graph
\textbf{8e}, where the HCMP
corresponds to the next highest local minimum for $L=2$ and $L=4$. Overall,
these observations further underline the importance of evaluating QAOA
performance based on the correct Max-Cut probabilities of individual minima
independently alongside their $\langle \hat{H}_C \rangle$ values.

The distributions of minima for the 3-regular graphs with varying $L$ differ
significantly from their complete graph counterparts (\textbf{Fig. 7}). We
observe that the convex hulls of the 3-regular graphs tend to take on a more
compact shape, with greater correlation between the 
$p(\ket{s})$ and $\langle \hat{H}_C \rangle$ values for the individual minima. However, for graphs
\textbf{8a}, \textbf{8b} and \textbf{8e} there is a notable absence of minima
with intermediate $p(\ket{s})$ and $\langle \hat{H}_C \rangle$ values,
particularly for higher $L$. This structure is also reflected in their
disconnectivity graphs (refer to \textbf{Appendix C} for the
disconnectivity graphs of \textbf{6a}-\textbf{8e}). Another major difference of
the 3-regular graphs is the much reduced energy differences between minima with
low proximity to the global minimum and their connected transition states,
producing more streamlined and single-funnelled disconnectivity graphs than for
$K_6$ and $K_8$. However, even though the energy landscapes of the 3-regular
graphs appear less complex and easier to navigate than 
their complete graph counterparts, the local minima in their energy
landscapes give rise to a larger range of $p(\ket{s})$. Hence a greater
proportion of local minima with high energies possess low Max-Cut
probabilities. This trend is captured by comparing $F$ values between the
3-regular graphs and the complete graphs $K_6$ and $K_8$, where the former
graphs typically have much higher $F$ values than the latter. 

Comparing $F$ values among the 3-regular graphs, we observe that graphs \textbf{6a} and
\textbf{6b} follow a smooth downward trend with increasing $L$, while the
8-vertex graphs tend to peak at $L=2$ and $L=3$ before decreasing, with the
exception of \textbf{8d}, which follows a similar trend of the 6-vertex graphs.
The performance of the 8-vertex graphs at $L=2$ and $L=3$ can be attributed to
an increase in the number of local minima with low Max-Cut probabilities that
outweigh the general improvement in HCMPs and low-lying minima, while for graph
\textbf{8d} this effect is reversed, with an increase in HCMPs
and minima with good probabilities. 
At $L=4$, a greater proportion of minima with high $p(\ket{s})$ appear for all 3-regular graphs, 
and we therefore recommend a minimum of $L=4$ when employing QAOA for these cases.
\begin{table}
\caption{\label{tab:4} Expectation thresholds $d_1$ (top value) and $d_2$ (bottom value) for graphs \textbf{6a}–\textbf{8e} of varying $L$.}
\begin{ruledtabular}
\begin{tabular}{cccc}
\textbf{Graph} & $L=2$ & $L=3$ & $L=4$ \\
\hline
\multirow{2}{*}{\centering \textbf{6a}}
      & 0.313189 & 0.546894 & 0.578616   \\
      & 0.643070 & 1.254700 & 1.342016   \\
\hline
\multirow{2}{*}{\centering \textbf{6b}}
      & 0.640787 & 1.453239 & 1.660319   \\
      & 0.951659 & 2.160089 & 2.470976   \\
\hline
\multirow{2}{*}{\centering \textbf{8a}}
      & - & 0.215689 & 0.338515  \\
      & - & 0.673507 & 1.251349  \\
\hline
\multirow{2}{*}{\centering \textbf{8b}}
      & -  & 0.250008 & 0.408135  \\
      & - & 0.886500 & 1.652028  \\
\hline
\multirow{2}{*}{\centering \textbf{8c}}
      & - & 0.046924 & 0.315151  \\
      & - & 0.183791 & 1.203264  \\
\hline
\multirow{2}{*}{\centering \textbf{8d}}
      & 0.052613 & 1.580115 & 1.857498   \\
      & 0.202901 & 2.457872 & 2.962362  \\
\hline
\multirow{2}{*}{\centering \textbf{8e}}
      & 0.096120  & 0.779749 & 0.645478   \\
      & 0.731083 & 1.578323 & 1.977320  \\
\end{tabular}
\end{ruledtabular}
\end{table}

Another factor that supports the choice of $L=4$ comes from the construction of
heuristic expectation thresholds that aim to identify minima with sufficiently
high $p(\ket{s})$ values. This analysis can be carried out by finding the intercepts of
the corresponding convex hulls with a suitable probability cutoff $p_{op}$.
For the 3-regular graphs we choose $p_{op}=0.5$ and define the difference in
$\langle \hat{H}_C \rangle$ values from the global minimum to the two
intercepts as the worst-case and best-case expectation cutoffs, $d_1$ and $d_2$,
respectively, where $d_1 < d_2$. We observe that the expectation thresholds
generally expand as $L$ increases (\textbf{Table IV}), with $d_1$ and $d_2$
attaining their highest values at $L=4$. For $L\leq{L_{min}}$ the widening and
stabilising of expectation thresholds is significant, along with the increase
in $M$ as $L$ increases. We see that a greater number of minima that possess a wider
range of $\langle \hat{H}_C \rangle$ values with a sufficiently high
$p(\ket{s})$ exist within the solution landscape for the QAOA ansatz.

\begin{figure}
\centering
\begin{tikzpicture}[scale=1.4, transform shape, node distance={15mm}, thick, main/.style = {draw, circle}] 
\node[main] (1) {$v_1$}; 
\node[main] (2) [above right of=1] {$v_2$}; 
\node[main] (3) [below right of=1] {$v_4$}; 
\node[main] (4) [above right of=3] {$v_3$};
\draw (1) -- node[midway, above right, sloped, pos=0.1] {3} (2); 
\draw (2) -- node[midway, above right, sloped, pos=0.1] {2} (4); 
\draw (3) -- node[midway, below left, sloped, pos=0.8] {1} (4); 
\draw (1) -- node[midway, below left, sloped, pos=0.9] {4} (3); 
\draw (2) -- node[midway, above right, pos=0.7] {$x$} (3); 
\end{tikzpicture}
\caption[Weighted graph $G_2$ to $G_5$ with variable central
weight]{Four-vertex weighted graph with a variable central weight $x$. The
graphs $(G_2, G_3, G_4, G_5)$ correspond to $x=(0,3,4,5)$ respectively.}
\label{fig:8}
\end{figure}
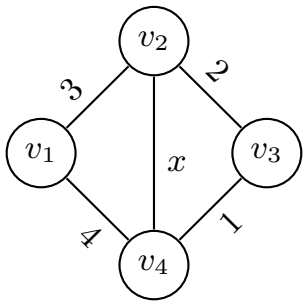

\begin{table}
\caption{\label{tab:5} Number of minima $M$ (top value), HCMPs (middle value)
and $F$ (bottom value) for graphs $G_2-G_5$ with varying $L$ obtained from
basin-hopping global optimisation. HCMPs with asterisks indicate that they
were obtained from the next highest local minimum instead of the global
minimum.}
\begin{ruledtabular}
\begin{tabular}{cccc}
\textbf{Graph} & $L=1$ & $L=2$ & $L=3$ \\
\hline
\multirow{3}{*}{\centering $G_2$}
      & 3    & 16    & 87   \\
      & 0.477824 & 0.866404 & 0.981615  \\
      & 0.594432 & 0.338167 & 0.163886   \\
\hline
\multirow{3}{*}{\centering $G_3$}
      & 7    & 109    & 1835   \\
      & 0.370522* & 0.784475 & 0.948704*  \\
      & 0.526539 & 0.300520 & 0.171897   \\
\hline
\multirow{3}{*}{\centering $G_4$}
      & 4    & 36    & 308   \\
      & 0.540426* & 0.894074 & 1.000000  \\
      & 0.461355 & 0.245864 & 0.111340   \\
\hline
\multirow{3}{*}{\centering $G_5$}
      & 9    & 183    & 3426   \\
      & 0.468114* & 0.881732* & 0.972620*  \\
      & 0.498733 & 0.328245 & 0.174754   \\
\end{tabular}
\end{ruledtabular}
\end{table}

\subsection{Competing QAOA Max-Cut solutions}
Finally, we explore competing Max-Cut solutions $\ket{\alpha\beta\alpha\beta}$
and $\ket{\alpha\alpha\beta\beta}$ for a series of four-vertex weighted graphs
with a common variable weight $x$, where $x=(0,3,4,5)$ correspond to the graphs
$(G_2,G_3,G_4,G_5)$, respectively (\textbf{Fig. 8}). These graphs open up two modes of
analysis: they allow comparison between $G_2$ with the more complex graphs
$G_3-G_5$, particularly with $G_3$, since both sets of graphs have
$\ket{s}=\ket{\alpha\beta\alpha\beta}$. Comparisons between $G_3-G_5$ can also
be carried out, since $G_5$ possesses a different correct Max-Cut solution of
$\ket{\alpha\alpha\beta\beta}$, while for $G_4$ both
$\ket{\alpha\beta\alpha\beta}$ and $\ket{\alpha\alpha\beta\beta}$ are correct
Max-Cut solutions. We will denote the alternative Max-Cut solution for a given
graph as $\ket{t}$, thus for $G_3$, $\ket{t}=\ket{\alpha\alpha\beta\beta}$; for
$G_5$, $\ket{t}=\ket{\alpha\beta\alpha\beta}$, and $G_4$ has no alternative
solution. 

We find that implementing QAOA for the weighted graphs $G_2-G_5$ is
more difficult than for the complete unweighted graph $K_4$, as their energy
landscapes are much more complex due to an increase in $M$ and a decrease in
their respective HCMPs (\textbf{Table V}). The disconnectivity graphs of
$G_2-G_5$ exhibit similar topological features to the 3-regular
unweighted graphs, possessing a well-funnelled organisation and minima
featuring a wide range of $\langle \hat{H}_C \rangle$ and Max-Cut probabilities
(see \textbf{Appendix D} for the disconnectivity graphs of $G_2-G_5$ coloured
based on $p(\ket{s})$, and \textbf{Appendix E} for the disconnectivity graphs
of $G_3$ and $G_5$ coloured based on $p(\ket{t})$). Unsurprisingly, $G_2$ has a
lower $M$ and thus a simpler landscape than the more strained graphs
$G_3-G_5$, although its collection of minima with modest Max-Cut probabilities
produces a comparatively high $F$ value up to $L=3$. In the range $G_3-G_5$,
it is interesting that even though $G_5$ with a different Max-Cut solution has
a more complex energy landscape than $G_3$, it yields a comparatively
higher HCMP, while $G_4$ exhibits the best QAOA performance as it has two
distinct Max-Cut solutions. The phenomenon where the HCMP arises for the next
highest local minimum rather than the global minimum was also observed for
$G_3-G_5$, especially for $G_5$ (refer to \textbf{Appendix B} for differences
in Max-Cut probabilities and $\langle \hat{H}_C \rangle$ values between the two
minima for these cases). All four graphs feature a monotonic decrease in $F$ as
$L$ increases. Hence, as for $K_5-K_8$, we recommend choosing $L=3$ in solving
the Max-Cut problem for $G_2-G_5$. Finally, it is noteworthy that for $G_4$ and
the $L=4$ ansatz, the number of minima greatly exceeds that of the $L_{min}=3$
ansatz, in contrast to the behaviour observed for $K_5$, highlighting the
general unpredictability of the energy landscape complexity after $L_{min}$.
\begin{table}
\caption{\label{tab:6} Expectation thresholds $d_1$ (top value) and $d_2$ (bottom value) for graphs $G_3-G_5$ of varying $L$.}
\begin{ruledtabular}
\begin{tabular}{ccc}
\textbf{Graph} & $L=2$ & $L=3$ \\
\hline
\multirow{2}{*}{\centering $G_3$}
      & 0.099242 & 0.433945   \\
      & 1.124694 & 2.008400   \\
\hline
\multirow{2}{*}{\centering $G_4$}
      & 0.489057 & 0.536238   \\
      & 1.447505 & 1.722258   \\
\hline
\multirow{2}{*}{\centering $G_5$}
      & 0.217591 & 0.459177   \\
      & 1.566966 & 1.896063   \\
\end{tabular}
\end{ruledtabular}
\end{table}
\begin{figure*}[t]
\centering
\includegraphics[width=17cm]{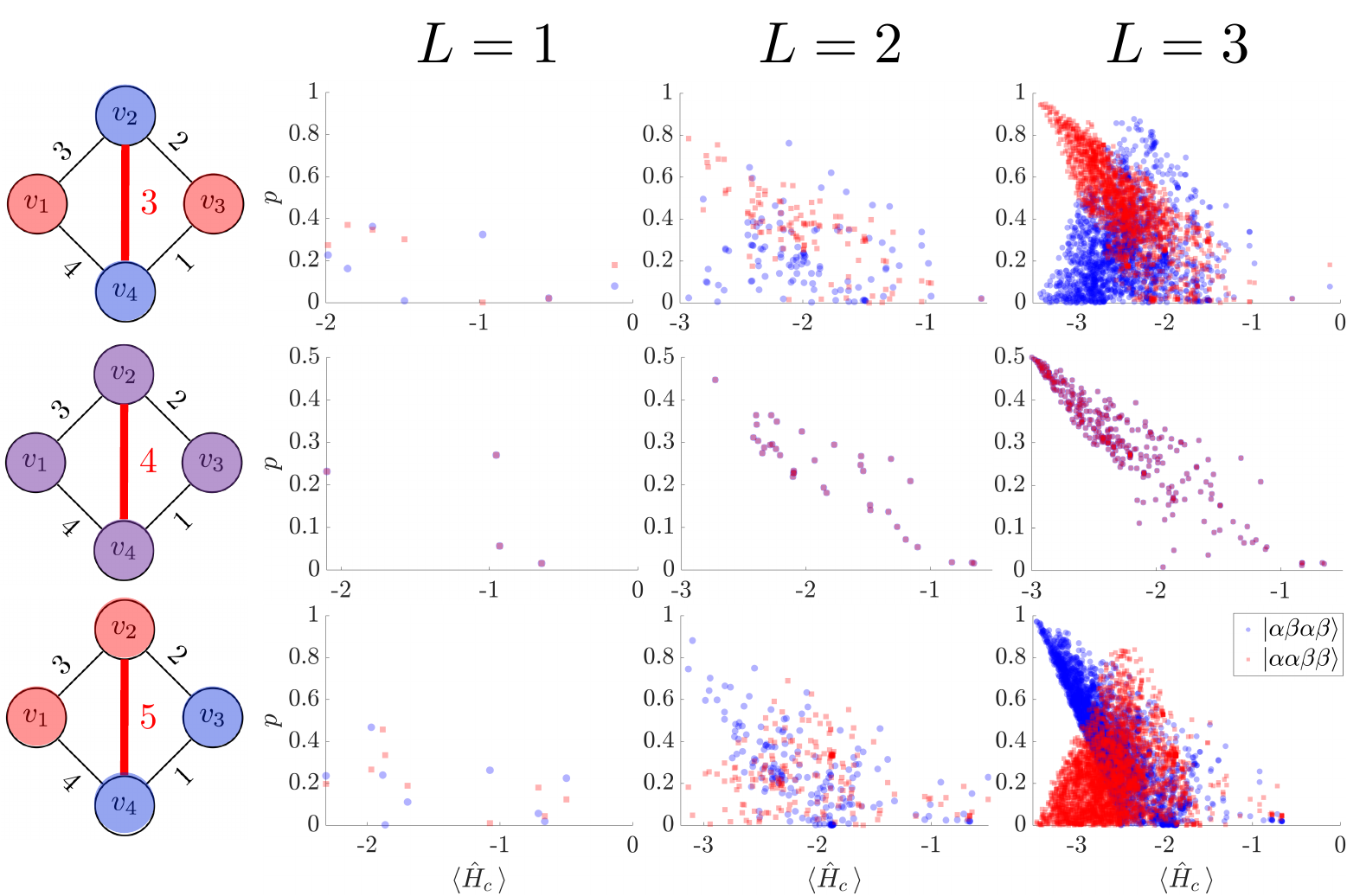}
\caption{\label{fig:9}Scatter plots of probabilities of competing Max-Cut
solutions $\ket{\alpha\beta\alpha\beta}$ (red square) and
$\ket{\alpha\alpha\beta\beta}$ (blue circle) against $\langle \hat{H}_C \rangle$
for connected minima of graphs $G_3$, $G_4$ and $G_5$ of varying $L$.}
\end{figure*}
For graphs $G_3-G_5$, the scatter plots of the $\langle \hat{H}_C \rangle$ values 
for connected minima and Max-Cut probabilities were used to construct both
the convex hulls of the correct Max-Cut solution $C_s$ and
the alternative solution $C_t$ (\textbf{Fig. 9}). We find that, similar to the
unweighted graphs, both $C_s$ and $C_t$ take on more definite shapes as they
become populated with more minima for increasing $L$. This trend allowed us to
investigate $d_1$ and $d_2$ by finding the intercepts of $C_s$ with the
horizontal line $p_{op}$, setting $p_{op}=0.5$ for $G_3$ and $G_5$, and
$p_{op}=0.25$ for $G_4$ (\textbf{Table VI}). Additionally, one may also use the
intercepts between $C_s$ and the left edge of $C_t$ as expectation and
threshold cut-offs to identify higher-quality minima with a high
probability of finding $\ket{s}$ and a low probability of obtaining $\ket{t}$.
(See \textbf{Appendix F} for more details.) As for the 3-regular graphs, the
divergence of $d_1$ and $d_2$ with increasing $L$ makes $L=3$ a good choice
for graphs $G_3-G_5$.

\section{Conclusion}
In this work we explore the solution landscapes of QAOA ans\"atze applied to a
variety of weighted and unweighted graphs by means of the energy landscapes
framework, using disconnectivity graphs to visualise their
topological features. We find that the corresponding landscapes are largely
funnelled, suggesting
that location of low-lying minima should not be particularly difficult.
Under practical conditions when simulating QAOA on a quantum device, the
optimisation regime is thus more likely to find a minimum close to the
global minimum with a good correct Max-Cut solution probability. Even under the
worst-case scenario where each experiment finds a different local minimum, so
long as the local minimum is sufficiently close to the global minimum, a
significant proportion of the number of shots per experiment will correspond to
the Max-Cut solution. This result further demonstrates the robustness of QAOA in solving
the Max-Cut problem. 

We have also developed a weighted average metric $F$ to
evaluate the performance of QAOA ans\"atze from their corresponding databases of
minima. This parameter allows one to choose a suitable number of circuit layers
that balances the likelihood of obtaining good solution probabilities from
local minima with an adequate circuit depth that minimises the impact from
quantum noise. 

Finally, we have established two ways in which expectation
thresholds can be established to determine the cut-off for minima with high
$p(\ket{s})$. The solution landscapes we have characterised suggest that
QAOA is a good VQA candidate to
demonstrate practical quantum advantage. In future work we plan to extend these results to quantum machine learning
(QML) algorithms, such as variational quantum classifiers (VQCs), which 
minimise a given cost function to classify data \cite{Jager2023}.

\section*{Software Availability}
The GMIN, OPTIM and PATHSAMPLE programs are available for
use under the Gnu General Public License. They can be downloaded from the 
Cambridge Landscape Database at www-wales.ch.cam.ac.uk.
\clearpage
\section*{Appendix A: Basin-hopping global optimisation with GMIN}
For each basin-hopping optimisation run, we performed 10,000 basin-hopping
steps for the unweighted graphs $K_3-K_8$, \textbf{6a}$-$\textbf{8e}, and for
the weighted graphs $G_2-G_5$, using the GMIN program for varying $L$. Each local
minimisation had a
minimum root-mean squared (RMS) gradient convergence criterion of
$1.0\times10^{-10}$ a.u., where the analytic gradients of the parameterised
rotation gates of the cost and mixer layers of the QAOA ansatz were evaluated
with the parameter-shift rule using \textbf{Eq. 7}.  To accept/reject basin-hopping steps we employed a Metropolis
criterion with a basin-hopping temperature of 1.0\, a.u. If the minimum at step $j$ has an expectation
value $\langle \hat{H}_C \rangle_{j}$ that is lower than the preceding
iteration, i.e. $\langle \hat{H}_C \rangle_{j}<\langle \hat{H}_C
\rangle_{j-1}$, $\langle \hat{H}_C \rangle_{j}$ then the corresponding angular
coordinates $\bm{\theta}_j$ are accepted and used for the next step.
If $\langle \hat{H}_C \rangle_{j} \geq \langle \hat{H}_C
\rangle_{j-1}$, then $\langle \hat{H}_C \rangle_{j}$ and $\bm{\theta}_j$ is
accepted with a probability of $\exp({-(\langle \hat{H}_C
\rangle_{j}-\langle \hat{H}_C \rangle_{j-1})/{k}T})$. 
Otherwise, the new minimum is rejected. 

Basin-hopping moves were proposed by 
random perturbations of up to 1.0\,rad for each angular coordinate in
$\bm{\theta}_j$.
At the end of each run, the collection of
minima that differ by at least $1.0\times10^{-9}$ a.u in their $\langle \hat{H}_C \rangle$ values were saved to provide a starting database for
construction of the energy landscape using the OPTIM and PATHSAMPLE programs.
\section*{Appendix B: Differences in Max-Cut probability and expectation values for non-global HCMP cases}
\begin{table}[h]
\caption{\label{tab:7} Expectation differences $\Delta\langle \hat{H}_C
\rangle$ and Max-Cut probability differences ${\Delta}p(\ket{s})$ between the
global and next highest local minima for non-global HCMP cases.}
\begin{ruledtabular}
\begin{tabular}{cccc}
\textbf{Graph} & $L$ & $\Delta\langle \hat{H}_C \rangle$ & ${\Delta}p(\ket{s})$ \\
\hline
\textbf{6a} & 3 & 0.037856 & 0.010121 \\
\hline
\textbf{8b} & 2 & 0.075696 & 0.040029 \\
\hline
\textbf{8e} & 2 & 0.046450 & 0.052876 \\
\hline
\textbf{8e} & 4 & 0.006337 & 0.009846 \\
\hline
$G_3$ & 1 & 0.127864 & 0.096306 \\
\hline
$G_3$ & 3 & 0.015904 & 0.003739 \\
\hline
$G_4$ & 1 & 1.142149 & 0.076922 \\
\hline
$G_5$ & 1 & 0.342188 & 0.231787 \\
\hline
$G_5$ & 2 & 0.036357 & 0.135824 \\
\hline
$G_5$ & 3 & 0.001111 & 0.001329 \\
\end{tabular}
\end{ruledtabular}
\end{table}
\textbf{Table VII} shows the expectation differences $\Delta\langle
\hat{H}_C \rangle = \langle \hat{H}_C \rangle_{min} - \langle \hat{H}_C
\rangle_\text{HCMP}$ and correct Max-Cut probability differences $\Delta
p(\ket{s}) = \text{HCMP} - p_\text{GM}(\ket{s})$ for cases where the next
highest local minimum has a Max-Cut probability greater than the global
minimum. In general, cases with a large circuit depth $L$ tend to have lower
$\Delta\langle \hat{H}_C \rangle$ and $\Delta p(\ket{s})$ as their HCMPs
approach the optimal value of 1. As mentioned in the main text this phenomenon
appears to occur sporadically, as exemplified by \textbf{8e} and $G_3$, where it
does not occur for the intermediate circuit depth of $L=3$. A notable case is
$G_5$, which features a higher Max-Cut probability in the next highest local
minimum for all sampled $L$. We propose to investigate this phenomenon in
future work to see if it arises
systematically for particular classes of connected graphs.
\clearpage
\onecolumngrid
\newpage
\section*{Appendix C: Disconnectivity graphs of the 3-regular graphs}
\begin{figure*}[h]
\centering
\includegraphics[width=17cm]{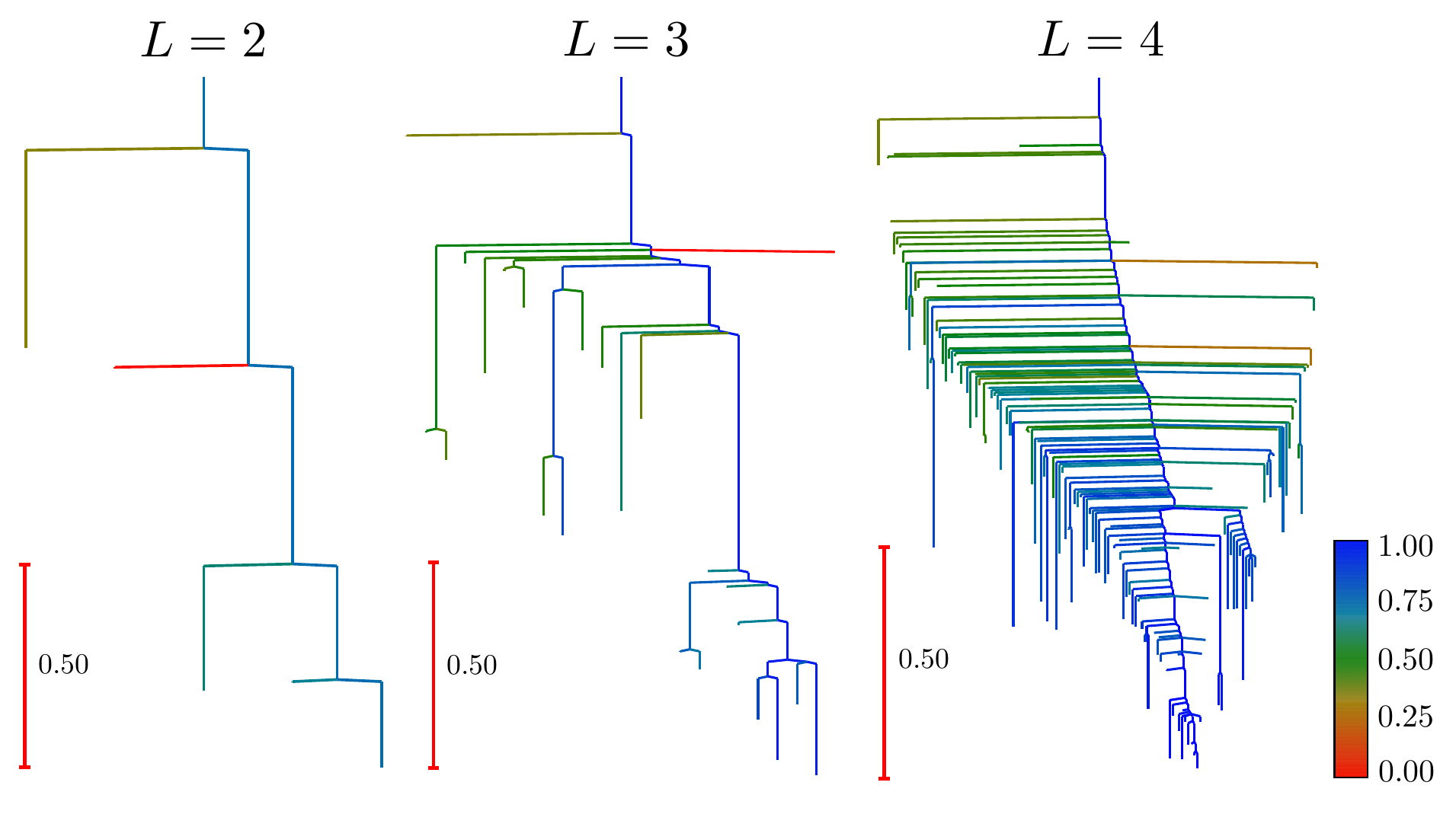}
\caption{\label{fig:10}Disconnectivity graphs for \textbf{6a} with varying circuit depth $L$.}
\vspace{2.0cm}
\includegraphics[width=17cm]{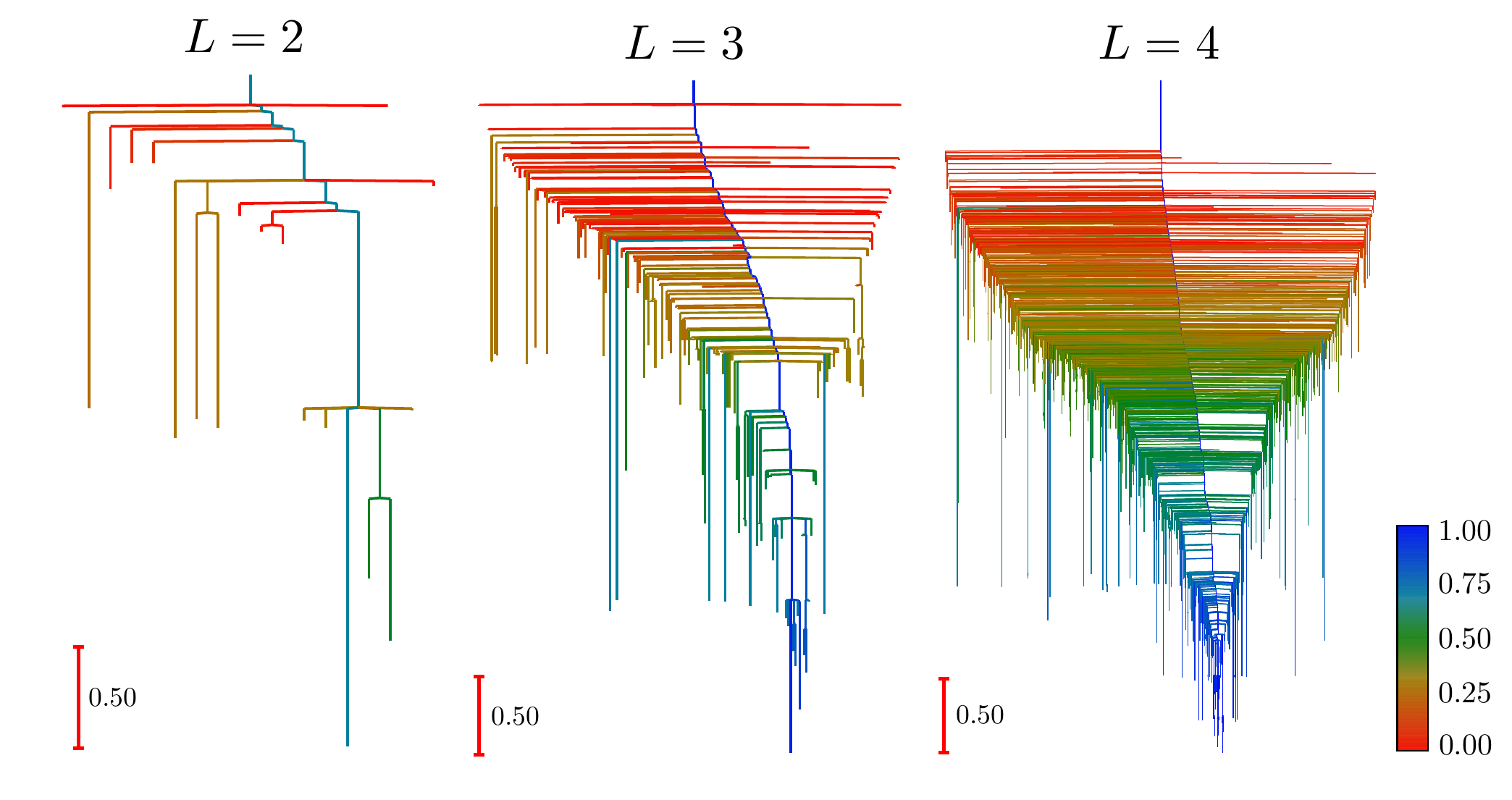}
\caption{\label{fig:11}Disconnectivity graphs for \textbf{6b} with varying circuit depth $L$.}
\end{figure*}

\begin{figure*}
\centering
\includegraphics[width=17cm]{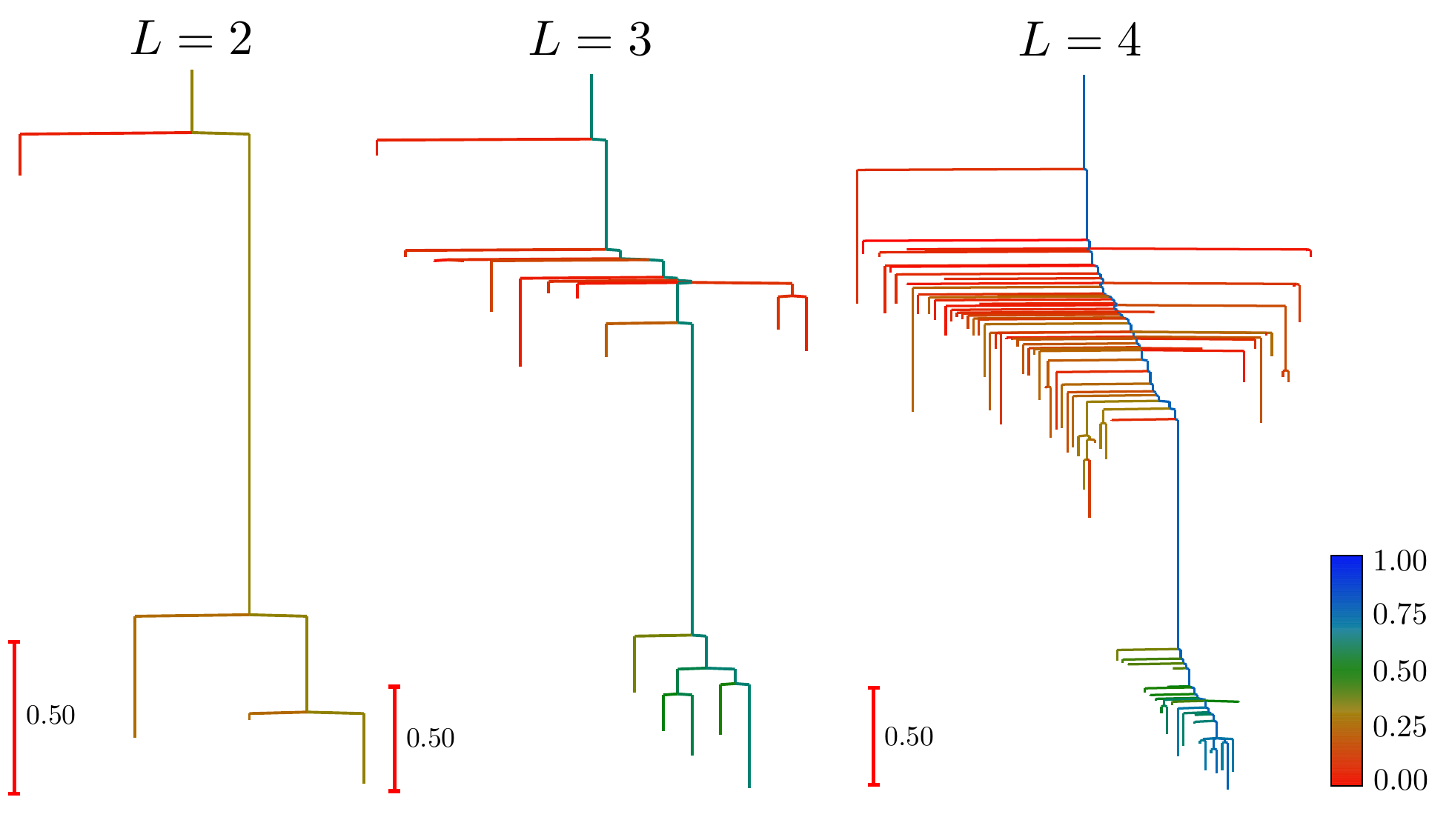}
\caption{\label{fig:12}Disconnectivity graphs for \textbf{8a} with varying circuit depth $L$.}
\centering
\vspace{2.0cm}
\includegraphics[width=17cm]{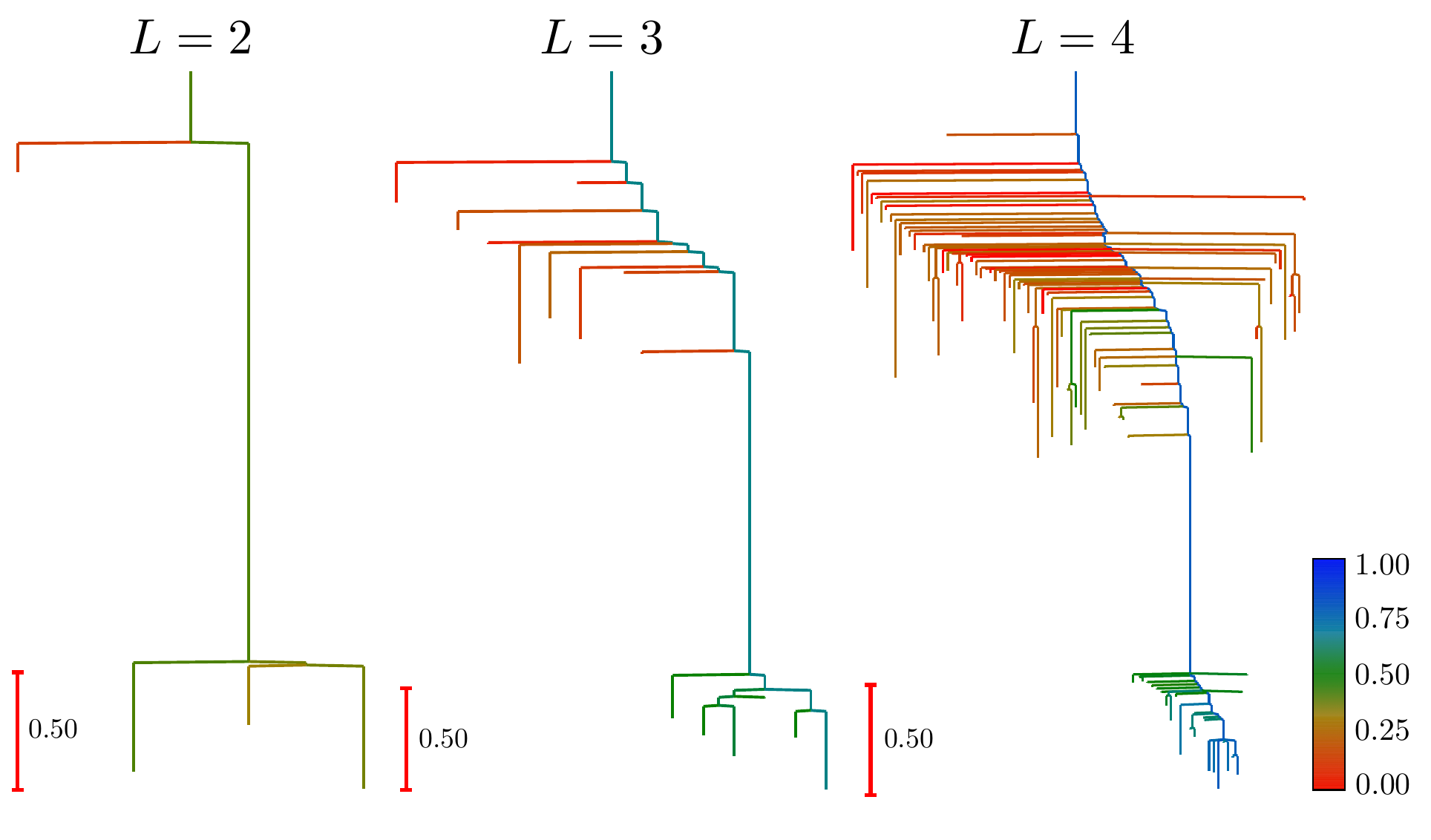}
\caption{\label{fig:13}Disconnectivity graphs for \textbf{8b} with varying circuit depth $L$.}
\end{figure*}

\begin{figure*}
\centering
\includegraphics[width=17cm]{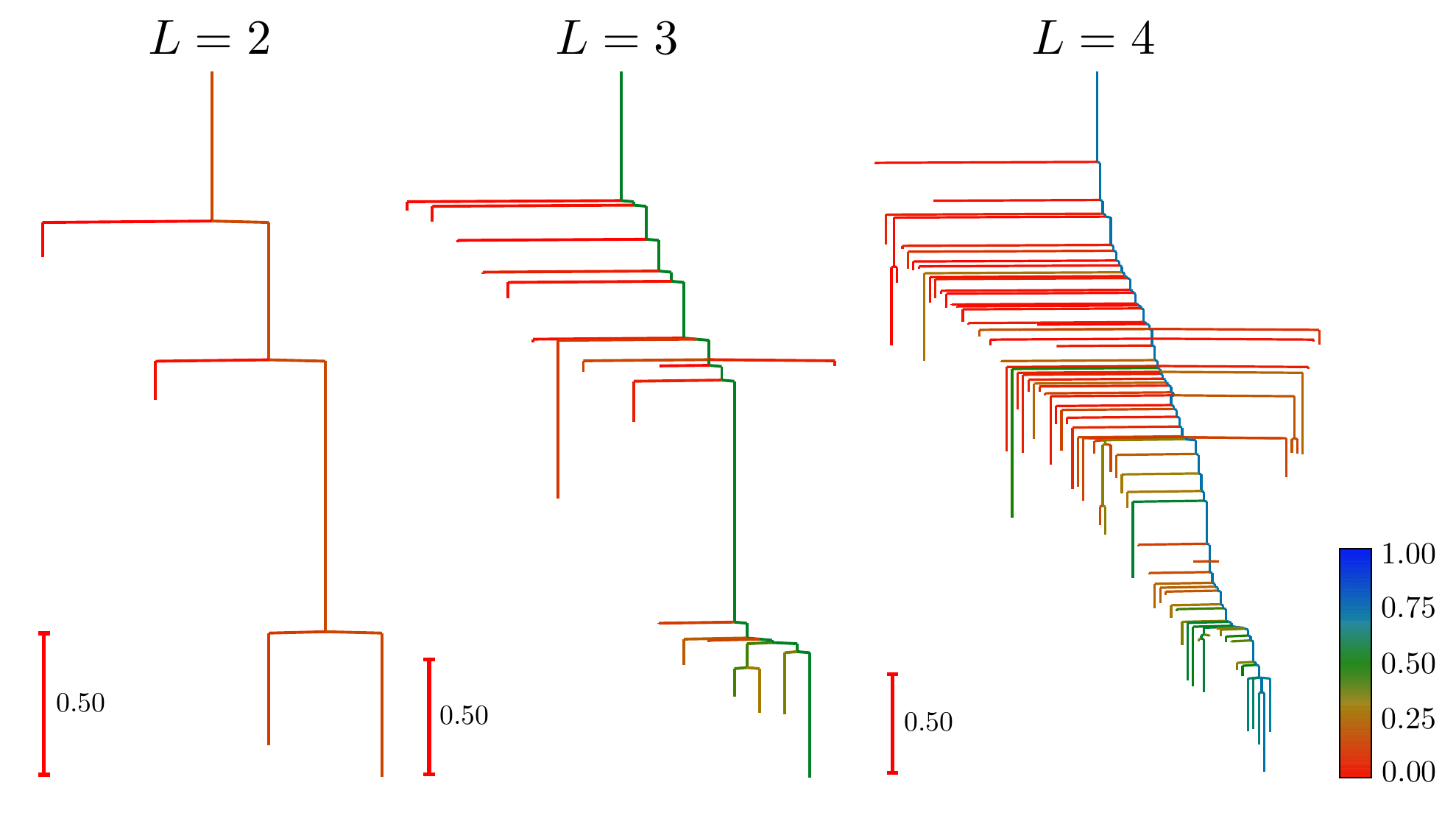}
\caption{\label{fig:14}Disconnectivity graphs for \textbf{8c} with varying circuit depth $L$.}
\centering
\vspace{2.0cm}
\includegraphics[width=17cm]{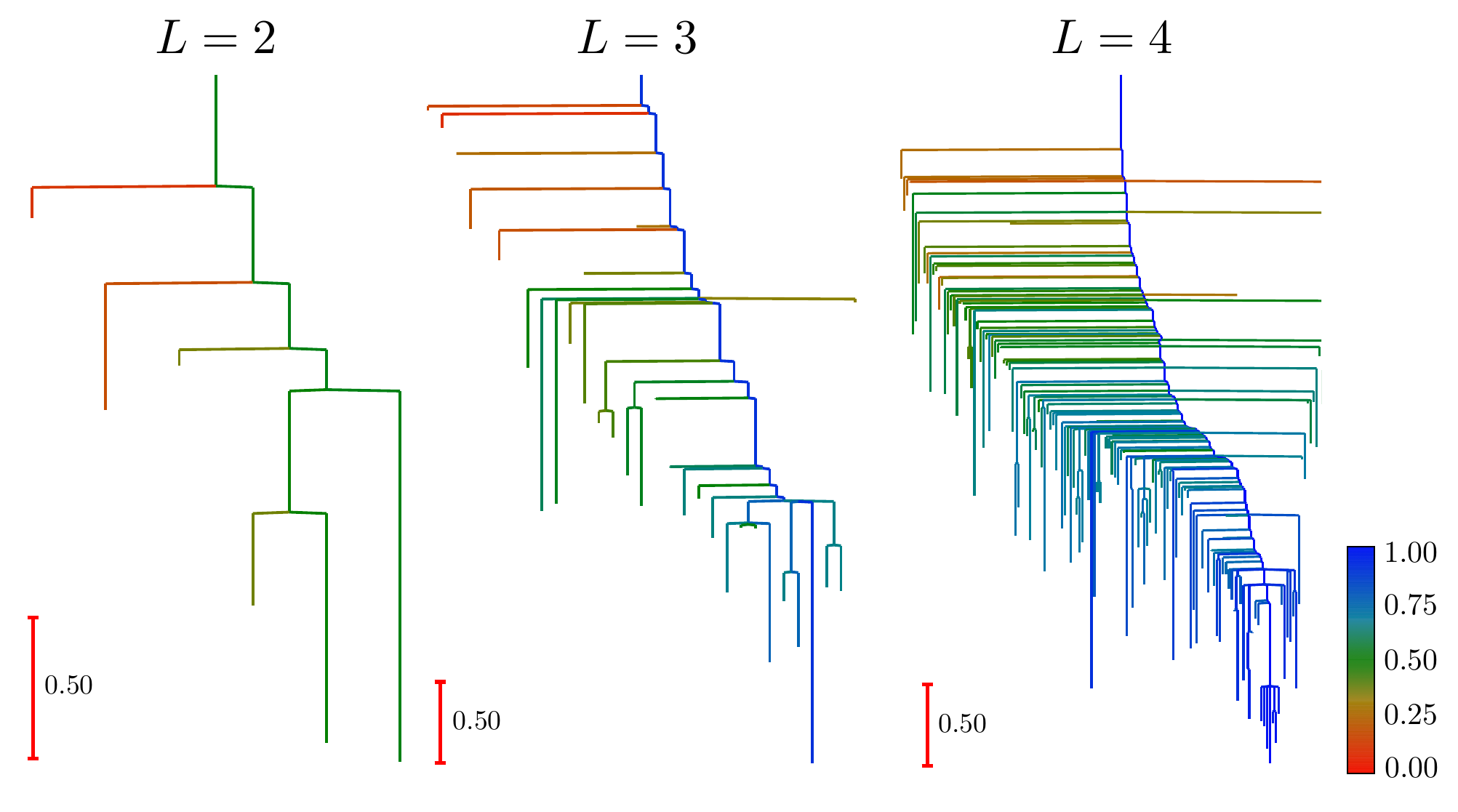}
\caption{\label{fig:15}Disconnectivity graphs for \textbf{8d} with varying circuit depth $L$.}
\end{figure*}

\begin{figure*}
\centering
\includegraphics[width=16.8cm]{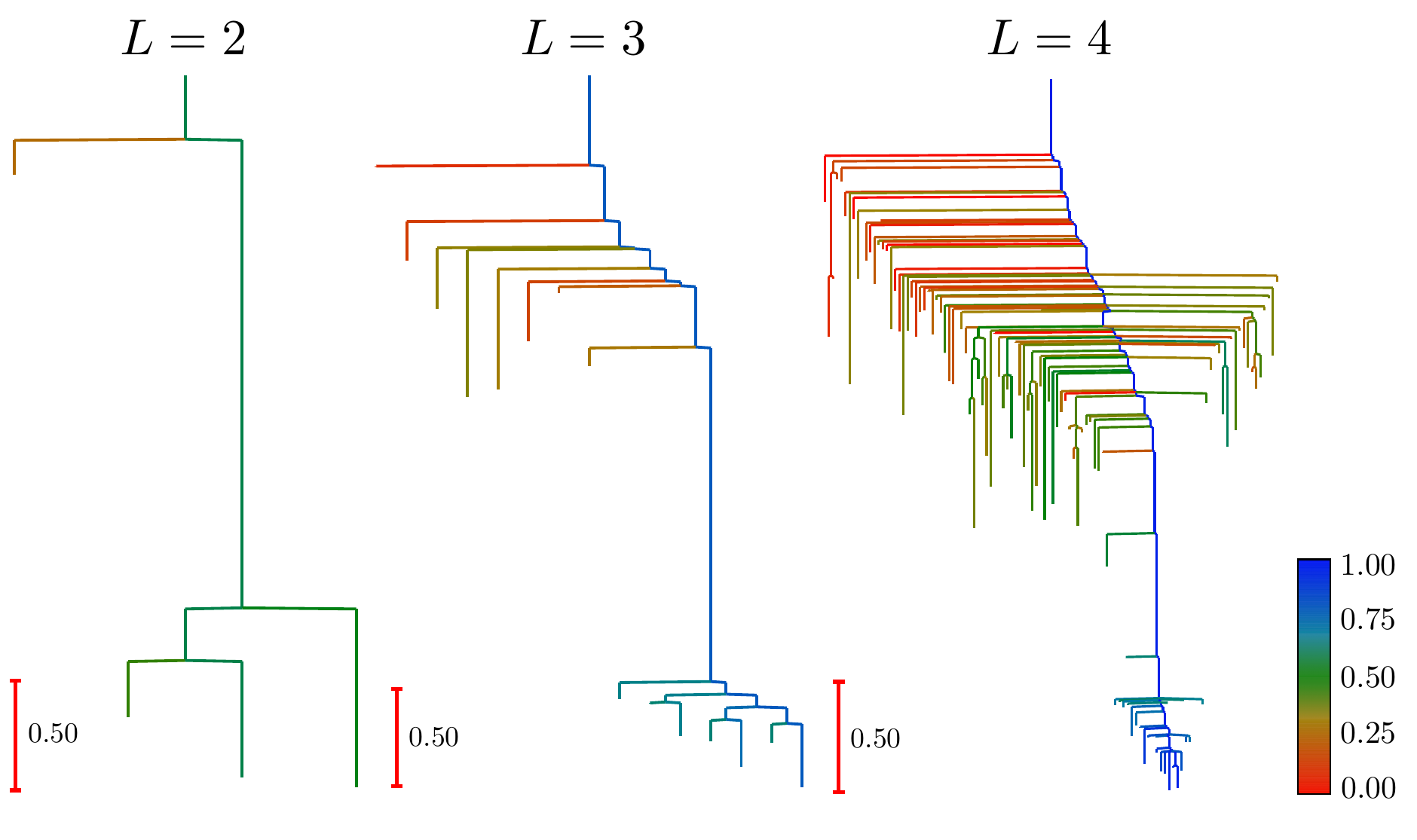}
\caption{\label{fig:16}Disconnectivity graphs for \textbf{8e} with varying circuit depth $L$.}
\end{figure*}
\clearpage
\section*{Appendix D: Disconnectivity graphs of the correct Max-Cut solutions for $G_2-G_5$}
\begin{figure*}[h]
\centering
\includegraphics[width=17cm]{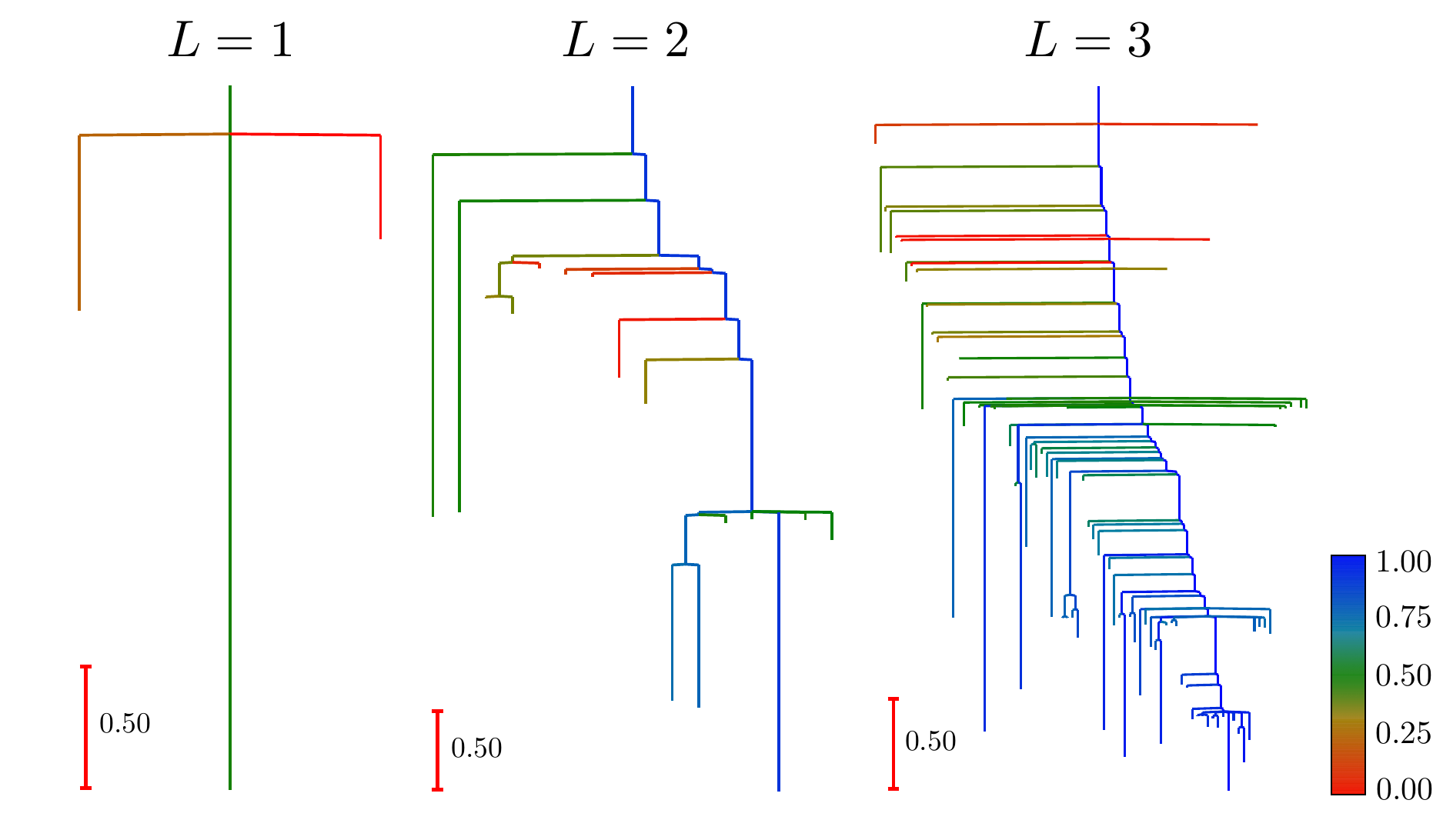}
\caption{\label{fig:17}Disconnectivity graphs for $G_2$ with varying circuit depth $L$. Minima are coloured based on the probability of obtaining the optimal Max-Cut state of $\ket{\alpha\beta\alpha\beta}$.}
\vspace{0.9cm}
\includegraphics[width=17cm]{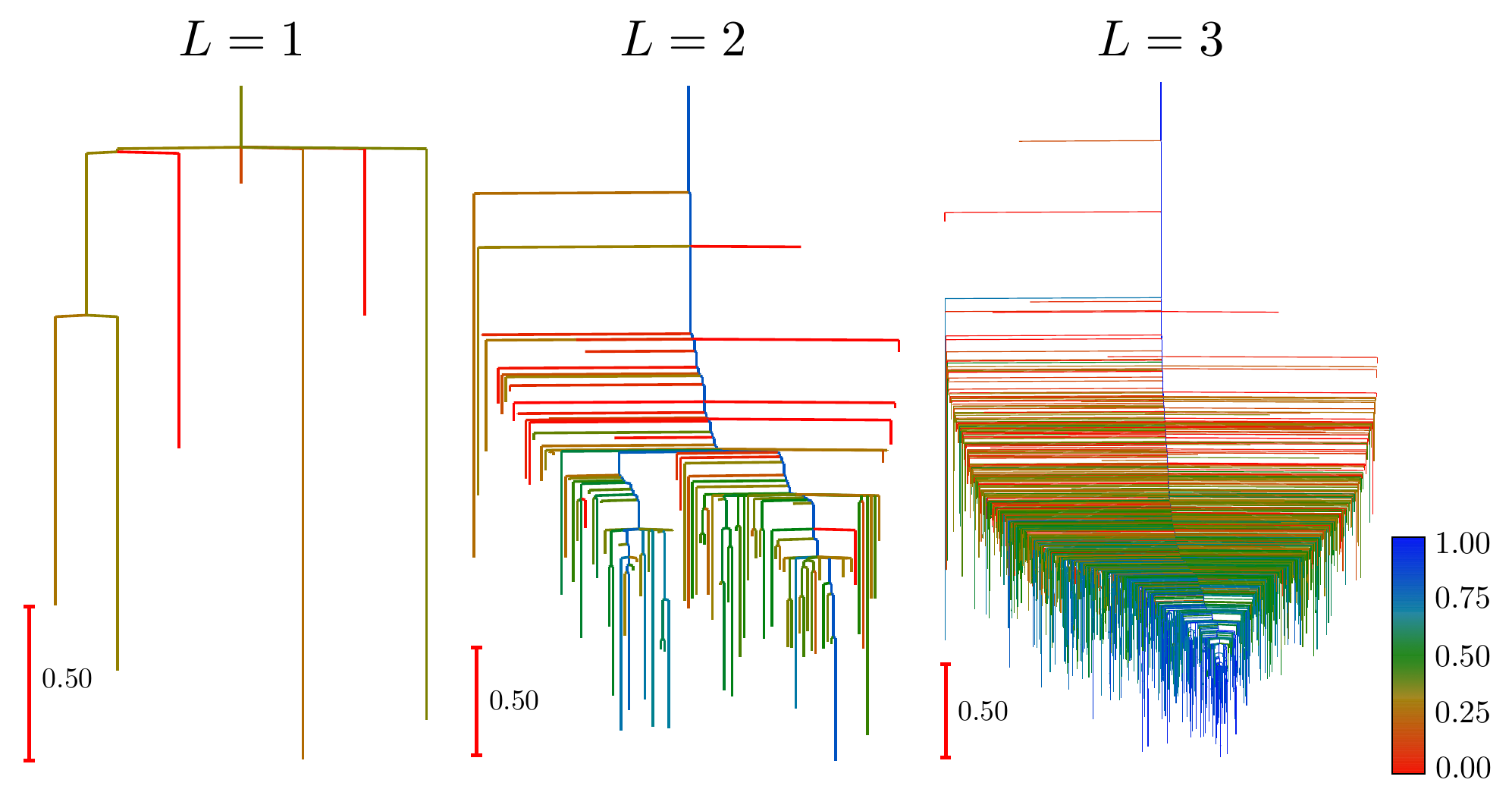}
\caption{\label{fig:18}Disconnectivity graphs for $G_3$ with varying circuit depth $L$. Minima are coloured based on the probability of obtaining the optimal Max-Cut state of $\ket{\alpha\beta\alpha\beta}$.}
\end{figure*}
\begin{figure*}
\centering
\includegraphics[width=17cm]{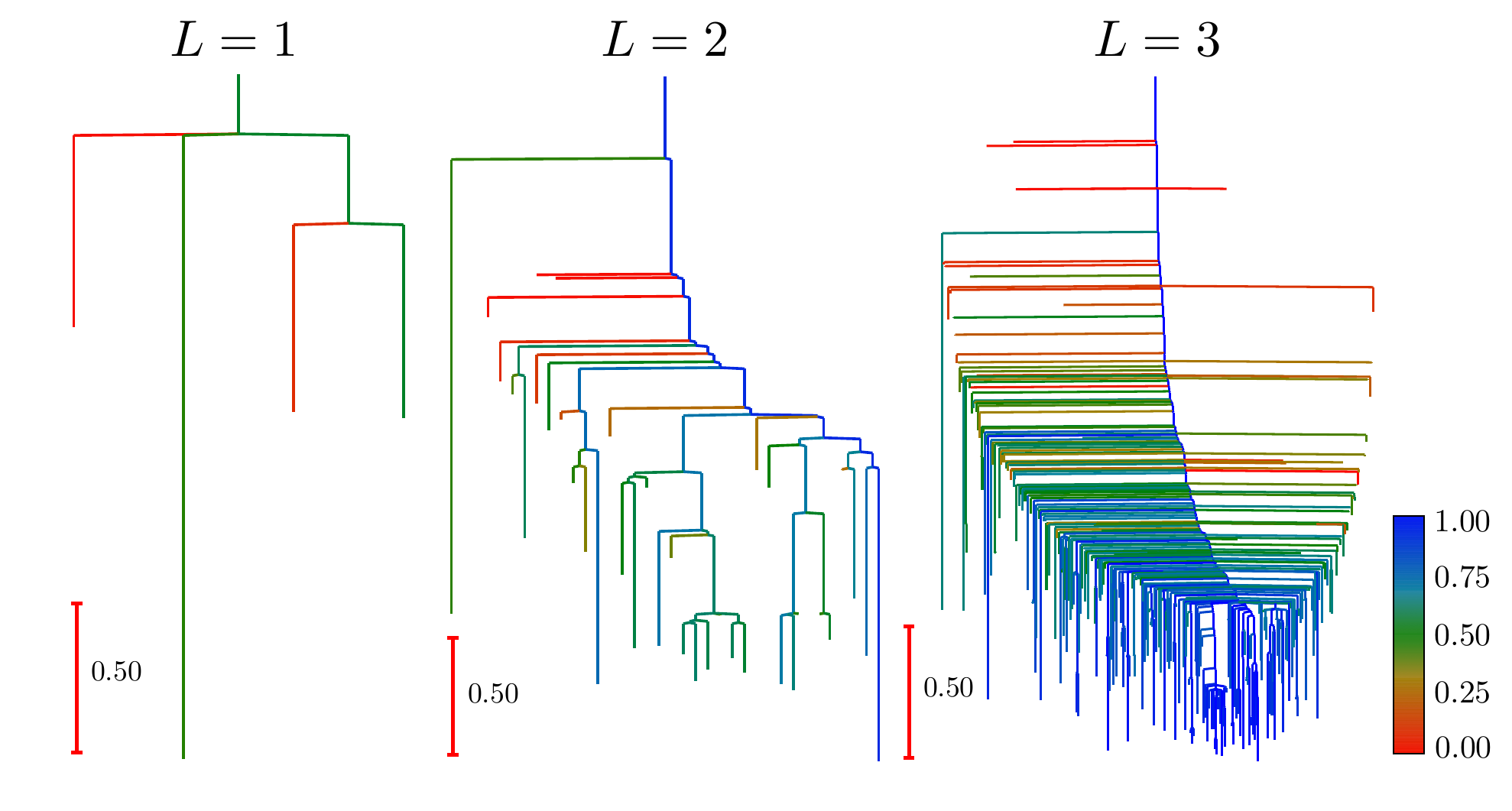}
\caption{\label{fig:19}Disconnectivity graphs for $G_4$ with varying circuit depth $L$. Minima are coloured based on the probability of obtaining the optimal Max-Cut states of $\ket{\alpha\beta\alpha\beta}$ and $\ket{\alpha\alpha\beta\beta}$.}
\vspace{2cm}
\includegraphics[width=17cm]{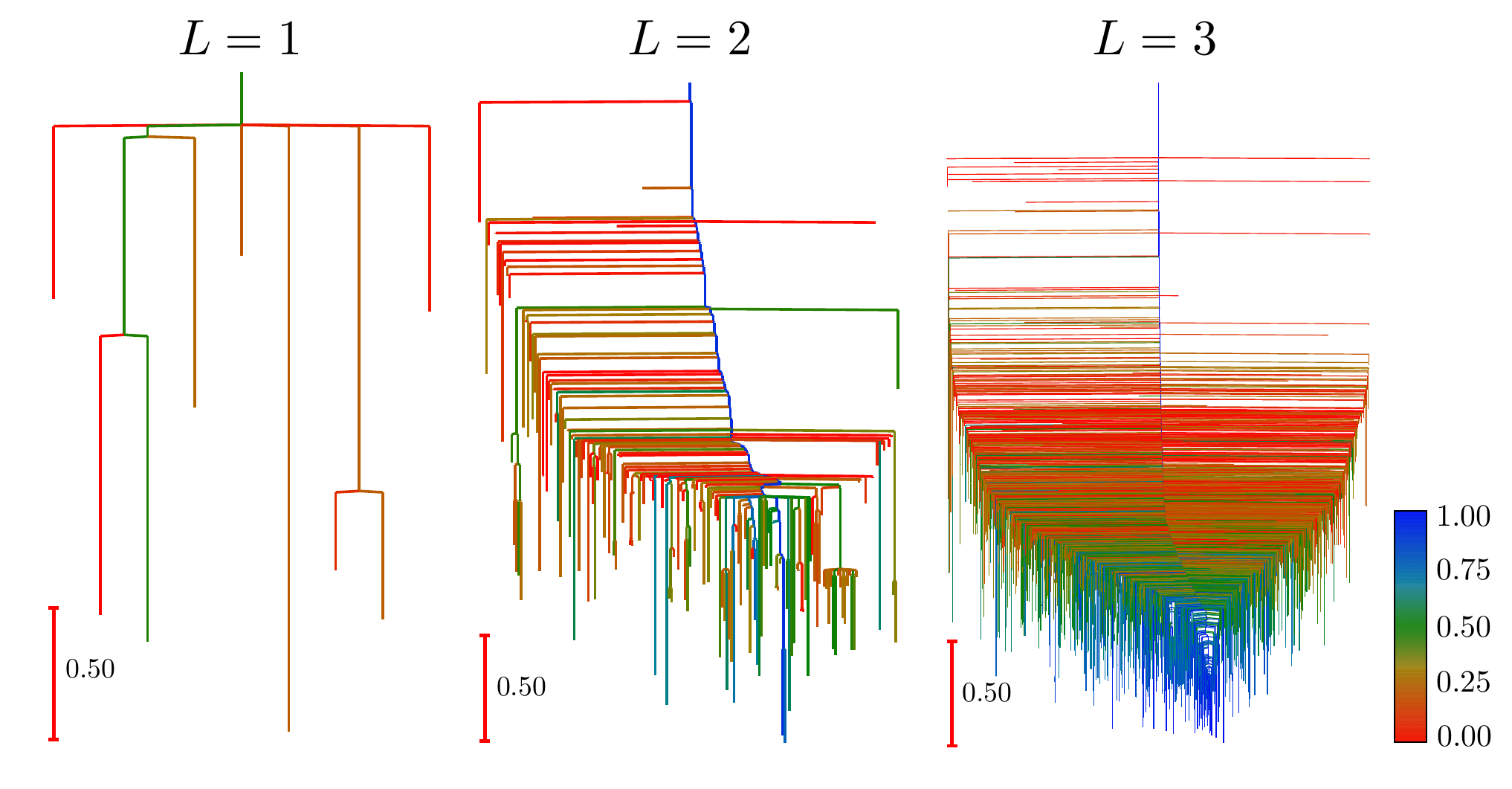}
\caption{\label{fig:20}Disconnectivity graphs for $G_5$ with varying circuit depth $L$. Minima are coloured based on the probability of obtaining the optimal Max-Cut state of $\ket{\alpha\alpha\beta\beta}$.}
\end{figure*}
\clearpage
\section*{Appendix E: Disconnectivity graphs of the alternative Max-Cut solutions for $G_3$ and $G_5$}
\begin{figure*}[h]
\includegraphics[width=17cm]{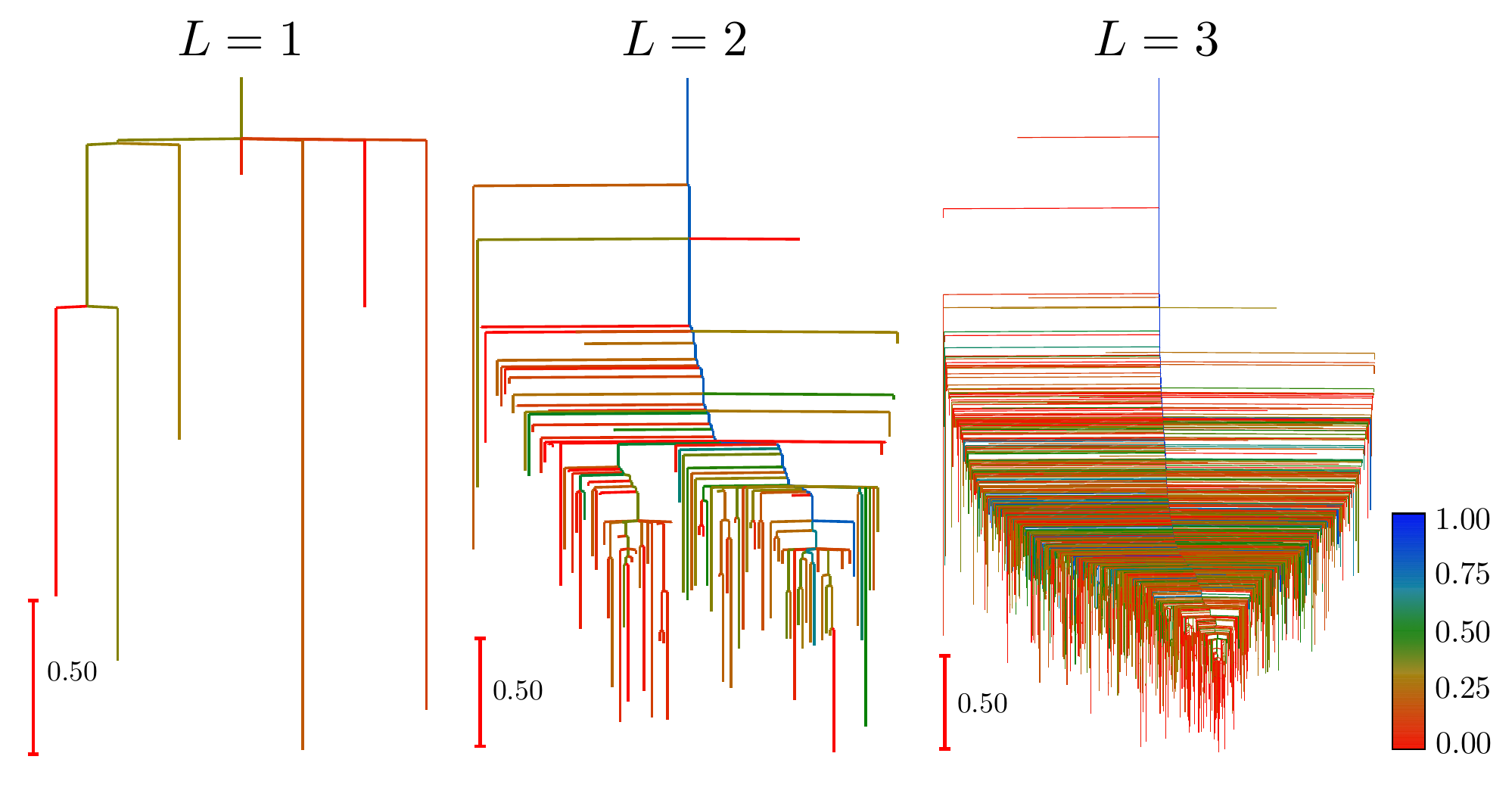}
\caption{\label{fig:21}Disconnectivity graphs for $G_3$ with varying circuit depth $L$. Minima are coloured based on the probability of obtaining the opposing Max-Cut state of $\ket{\alpha\alpha\beta\beta}$.}
\vspace{1.4cm}
\includegraphics[width=17cm]{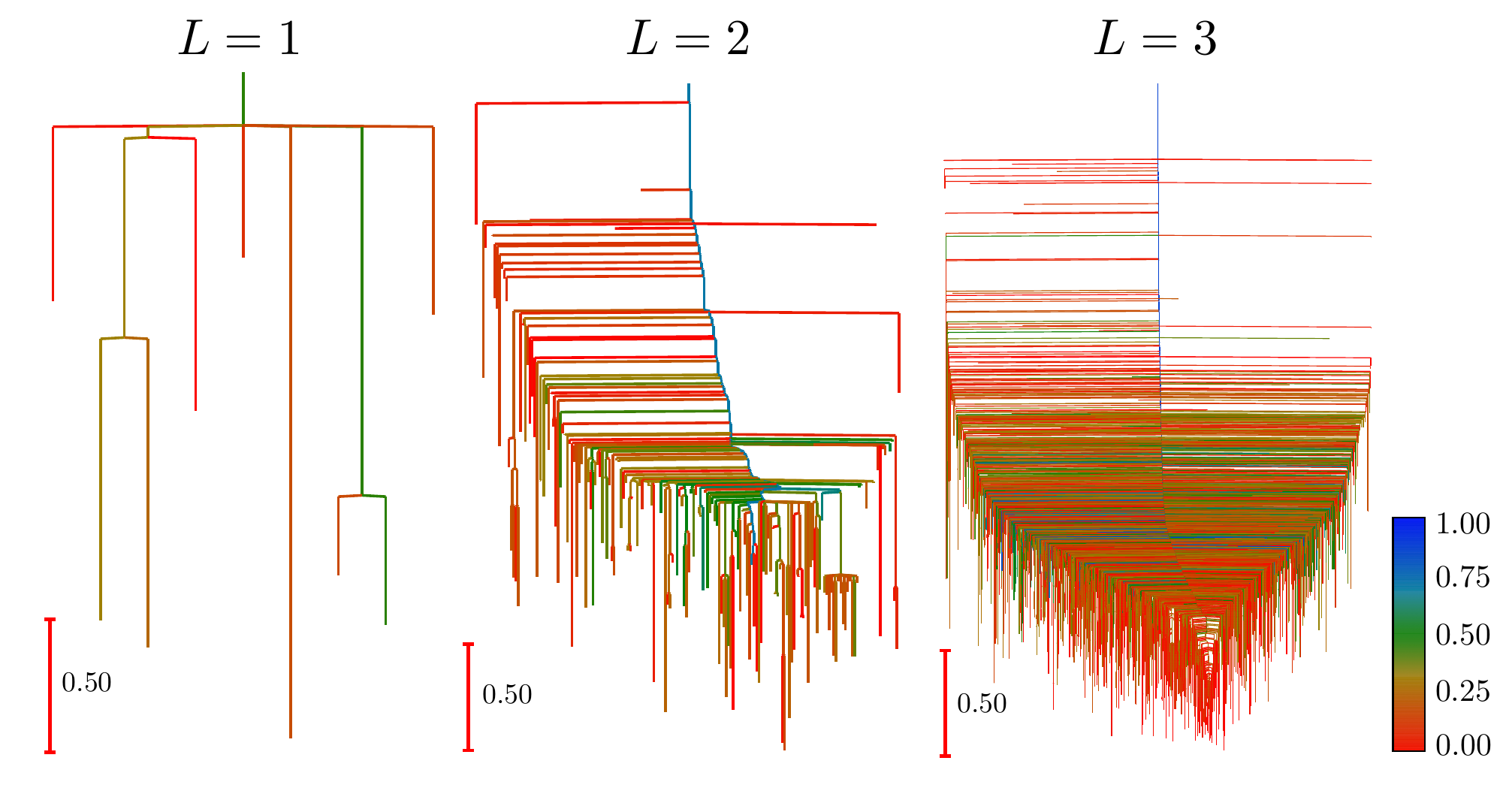}
\caption{\label{fig:22}Disconnectivity graphs for $G_5$ with varying circuit depth $L$. Minima are coloured based on the probability of obtaining the opposing Max-Cut state of $\ket{\alpha\beta\alpha\beta}$.}
\end{figure*}
\clearpage
\twocolumngrid
\section*{Appendix F: Expectation and probability thresholds from the convex hull intercepts for $G_3$ and $G_5$}
\begin{table}[ht]
\caption{\label{tab:8} Expectation thresholds $d_3$ (top value) and $d_4$ (bottom value) for graphs $G_3$ and $G_5$ of varying $L$, where $d_3 < d_4$.}
\begin{ruledtabular}
\begin{tabular}{ccc}
\textbf{Graph} & $L=2$ & $L=3$ \\
\hline
\multirow{2}{*}{\centering $G_3$}
      & 0.111125    & 0.439036     \\
      & 0.546385    & 0.794808     \\
      \hline
\multirow{2}{*}{\centering $G_5$}
      & 0.329259    & 0.476747     \\
      & 0.856478    & 0.819369     \\
\end{tabular}
\end{ruledtabular}
\end{table}
\begin{table}[ht]
\caption{\label{tab:9} Max-Cut probability thresholds $p_1$ (top value) and $p_2$ (bottom value) for graphs $G_3$ and $G_5$ of varying $L$, corresponding to the expectation thresholds $d_3$ and $d_4$ respectively.}
\begin{ruledtabular}
\begin{tabular}{ccc}
\textbf{Graph} & $L=2$ & $L=3$ \\
\hline
\multirow{2}{*}{\centering $G_3$}
      & 0.465937    & 0.495366     \\
      & 0.665681    & 0.799744     \\
      \hline
\multirow{2}{*}{\centering $G_5$}
      & 0.373801    & 0.482335     \\
      & 0.680284    & 0.790380     \\
\end{tabular}
\end{ruledtabular}
\end{table}
\textbf{Tables VIII} and \textbf{IX} summarise the two intercepts $(\langle
\hat{H}_C \rangle_{min} + d_3, p_1)$ and $(\langle \hat{H}_C \rangle_{min} +
d_4, p_2)$ of $C_s$ with the left edge of the triangular convex hull $C_t$ for
graphs $G_3$ and $G_5$, as shown in \textbf{Fig. 9}. This summary provides an
alternative method to establish expectation cut-offs if additional information
for alternative Max-Cut states is available. As for the expectation thresholds
$d_1$ and $d_2$ in \textbf{Table VI}, $d_3$ and $d_4$ also increase with
increasing $L$, which further justifies the choice of $L=3$ in refining minima
within $C_s$ that possess both good correct Max-Cut probabilities and low
opposing Max-Cut probabilities. The probability cut-off of $p_{op}=0.5$ used in
\textbf{Table VI} also appears to be situated between the probability thresholds
$p_1$ and $p_2$, suggesting that this choice of $p_{op}$ is optimal.

\clearpage
\bibliography{qaoa}

\end{document}